\documentclass[aps,notitlepage]{revtex4-1}
\usepackage[utf8]{inputenc}
\usepackage{amsmath,amssymb}
\usepackage{graphicx,float}
\usepackage{dsfont}
\usepackage{multirow}
\usepackage{xfrac}
\usepackage{resizegather}
\usepackage{standalone}

\makeatletter
\usepackage[bottom]{footmisc}
\usepackage{nicefrac}

\newcommand{\iu}{\mathrm{i}\mkern1mu}

\usepackage{xr}
\usepackage[a4paper, total={6in, 9in}]{geometry}

\usepackage{color}

\newcommand{\1}{\mathds{1}}

\let\oldtop\top
\renewcommand{\top}{\oldtop\!}

\begin{document}

\title{How does bond percolation happen in coloured networks?}
\author{Ivan Kryven} 
\email{i.kryven@uva.nl}
\affiliation{Van 't Hoff Institute for Molecular Sciences, University of Amsterdam, Science Park 904, 1098 XH Amsterdam, The Netherlands} 

\begin{abstract}
Percolation in complex networks is viewed as both: a process that mimics network degradation and a tool that reveals peculiarities of the underlying network structure. 
During the course of percolation, networks undergo non-trivial transformations that include a phase transition in the connectivity, and in some special cases, multiple phase transitions. 
Here we establish a generic analytic theory that describes how structure and sizes of all connected components in the network are affected by simple and colour-dependant bond percolations. This theory predicts all locations where the phase transitions take place, existence of wide critical windows that do not vanish in the thermodynamic limit, and a peculiar phenomenon of colour switching that occurs in small connected components.
These results may be used to design percolation-like processes with desired properties, optimise network response to percolation, and detect subtle signals that provide an early warning of a network collapse.
\end{abstract}
\maketitle

\section*{}
One of the richest tools for exploring properties of complex networks is probing these networks by randomly removing links.
This process is called percolation, and on many occasions percolation has  shaped our understanding of physical and socio-economic phenomena\cite{DSouza2015}. Naturally, percolation is related to network resilience. This connection was exploited, for instance, in studies on communication, transportation, and supply networks  \cite{Duncan2000,onnela2007,osat2017,radicchi2015}.
Percolation has defined the modern view on disease epidemics \cite{Duncan2000,newman2002b,davis2008,zuzek2015,allard2017}, as well as on other spreading processes   \cite{Radicchi2009,roukny2013,Manlio2016}. 
Percolation is instrumental in material science, where the gelation  \cite{coniglio1979,Kryven2016a,jaspers2017,kryven2018} and jamming  \cite{family2012,papadopoulos2018} have been both connected to this process.

It is known that during percolation, complex networks undergo a series of non-trivial transformations that include splitting of connected components and criticality in the global connectivity  \cite{newman2001,baxter2012,bianconi2014,radicchi2015,Gleeson2008}.
Remarkably, even the simplest models that define a network solely by its degree distribution, the configuration models, do also feature these phenomena. 
In many studies, this observation justified the use of the configuration model as the null-model, which one would compare against any real network, in order to reveal presence of trends that are, or are not, explainable by the degree distribution alone.

In the edge-coloured configuration model, the edges are categorised into different groups or colours. This distinction leads to a more realistic representation of complex networks. Indeed, most of real networks do feature different types of interactions:
 chemical bonds,  communication channels, social interactions between infected individuals, among other examples, all require a partition of the edges into discrete categories  \cite{kivela2014,hackett2016,allard2017,klosik2017}.
Furthermore, a whole new world of possibilities unfolds when the number of colours is virtually unlimited:
different edge colours  may encode all combinations of one interaction occurring between different types of nodes, 
as was proposed in the assortative mixing model  \cite{newman2002a}, or one type of interaction with discrete strengths  \cite{mastrandrea2014}.
In modular networks, interactions within each community can be represented with a unique colour.
In contrast to unicoloured networks  \cite{newman2010},  less is known about how percolation happens in coloured networks. 
For instance, presence of multiple phase transitions has been reported  \cite{baxter2012,bianconi2014,hackett2016}, but no complete theory could explain their nature nor predict locations where these phase transitions occur. It has been observed that negative correlation in the degree distribution leads to multiple phase transitions \cite{hackett2016}.

This paper demonstrates that in order to explain the percolation in coloured networks, one has to study not only the giant component but also the rest of the system.
In fact, the tail asymptote for the sizes of connected components already covers sufficient amount of information. 
By building upon this result, we establish universal criteria of phase transitions for two percolation processes: simple bond percolation, which removes every edge with equal probability, and the colour-dependent bond percolation, in which edges of different colours have different probability to be removed.
This theory detects multiple phase transitions, if such occur, and in the case of colour-dependant percolation, the theory yields a manifold containing all critical points. 
Surprisingly, the theory predicts existence of wide critical windows that do not vanish in infinite systems and colour switching that occurs in small connected components during percolation.

\section{Results}

\subsection{The $N$-colour asymptotic theory}
 Newman et al. observed that in a unicoloured network defined by the degree distribution, the size distribution of connected components can be found numerically  \cite{newman2001}.
 His ideas have been later formalised into an analytical theory in our previous work  \cite{kryven2017a} (see SI~1 for a brief summary).
The current work takes this analytical concept further by considering a network with an arbitrary number of colours.
Suppose every edge has one of $N$ colours, so that a randomly chosen node bears $k_1$ edges of colour one, $k_2$ edges of colour two,  and so on. A state of a node
is parametrised by a vector of colour numbers $\boldsymbol k = (k_1,\dots,k_N)$. Let $u(\boldsymbol k)$ be the probability that a randomly chosen node has state $\boldsymbol k$. 
In this paper, we often refer not to $u(\mathbf k)$ directly, but to its parameters $\boldsymbol \mu_0,\boldsymbol M,$ and $ \boldsymbol T_i$ that are defined via expectations of $u(\mathbf k)$:
\begin{equation}\label{eq:M}
\boldsymbol \mu_0  =(\mathbb E[k_1],\mathbb E[k_2],\dots,\mathbb E[k_N])^\top,
\end{equation}
\begin{equation}\label{eq:M}
M_{i,j}  = \frac{ \mathbb E[ k_i k_j ]}{\mathbb E[ k_j]}-\delta_{i,j},\; i,j=1,\dots,N,
\end{equation}
\begin{equation}\label{eq:T}
(\boldsymbol T_i)_{j,l} = \frac{\mathbb E[ k_i k_j  k_l ]}{\mathbb E[ k_i]}
- \frac{\mathbb E[k_i k_j ]\mathbb E[k_i k_l ]}{\mathbb E[ k_i]^2},\; i,j,l=1,\dots,N.
\end{equation}
The probabilistic interpretation of $\boldsymbol \mu_0,\boldsymbol M, \boldsymbol T_i$ and the concept of the expected value is explained in the Methods.
In our previous paper  \cite{kryven2017b},  the formal expression for the  size distribution of connected components $w(n)$ was derived by applying Joyal's theory of species,
\begin{equation}\label{eqA1:multiplex}
w(n)  = \sum_{\substack{k_1+\dots+k_N=n+1\\k_i \geq 0}} \left(\boldsymbol D* u *  u_1^{*k_1}*\dots * u_N^{*k_N} \right)(\mathbf k),
\end{equation}
where operations $f*g$ and $f^{*k}$ denote, respectively, the $N$-dimensional convolution product and the convolution power  \cite{kryven2017b}, whereas $\boldsymbol D$ is the auxiliary function  defined in SI~2.
Although the computations of equation \eqref{eqA1:multiplex} require large complexity of $O(n^N \log n)$, it is possible to write the tail asymptote for arbitrary number of colours $N$:
\begin{equation}\label{eq:simple_asm}
w_\infty(n)=C_1 n^{-\sfrac{3}{2} } e^{- C_2 n}, \; n\gg1.
\end{equation}
Here coefficients $C_1>0,C_2\geq 0$ are defined in terms of $\boldsymbol \mu_0, \;\boldsymbol M$ and $\boldsymbol T_i.$ The Methods provide the expressions for these coefficients, and  SI~3 contains the derivation of equation \eqref{eq:simple_asm}. 
Coefficient $C_2$ is especially important as it defines how fast the size distribution decreases:
$C_2\gg0$ implies exponentially rapid decrease; $C_2=0,$ triggers  a scale-free behaviour, and close-to-zero values of $C_2$ are associated with transient scale-free behaviour that is eventually overrun by an exponentially fast decrease. 
The universality of $-\frac{3}{2}$ constant has been noticed by Newman et al. in unicoloured networks   \cite{newman2001}.
Remarkably, the exponent $-\frac{3}{2}$ also holds universally for arbitrary number of colours $N$ and an arbitrary multi-degree distribution that features finite mixed moments $\mathbb E[ k_i k_j  k_l ]$.

In the configuration models, a connected component of size $n$ has a tree-like structure, and therefore, 
contains $n-1$ edges.
Let $v_i$ denote the numbers of $i-$coloured edges that are present in a component of size $n\gg1$.
By normalising with the total number of edges $f_i:=\frac{v_i}{n-1}$, one obtains the respective colour fractions.
 The probability of a particular configuration of colour fractions $\mathbb P[f_1,f_2,\dots,f_N]$ is the Gaussian function with independent of $n$ mean $\mathbf m$ and covariance matrix $\frac{1}{n}\boldsymbol \Sigma$ that vanishes when the component size tends to infinity. The expressions for $\mathbf m$ and $\boldsymbol \Sigma$ are given in terms of $\boldsymbol M$ and $\boldsymbol T_i$ in the Methods, and the necessary derivations can be found in SI 7. 

\subsection{Criticality and simple bond percolation}
There exists a class of degree distributions for which $C_2=0$ and therefore the asymptote \eqref{eq:simple_asm} is  scale-free.
We call these degree distributions critical.
As shown in SI~4, the critical degree distribution can be identified by testing with a following criterion: $C_2 =0$ if and only if 
\begin{equation}\label{eq:criterion}
\mathbf v \in \ker [ \boldsymbol M - \boldsymbol I],\; \text{and } \frac{\mathbf v}{| \mathbf  v|} > 0.
\end{equation}
Before testing with criterion \eqref{eq:criterion}, it is useful to check first if $\det[ \boldsymbol M-\boldsymbol I ]=0,$ which is only a necessary condition for criticality but not a sufficient one. 
Equation \eqref{eq:criterion} generalises many known relations, for instance,
Molloy-Reed criterion, equation 33 in   \cite{newman2001};
the criterion for the multiplex giant component, equation 11 in  \cite{allard2015}, 
or the message passing criticality, equation 27 in Ref. \cite{bianconi2017}.

Now, let us investigate a simple process that continuously changes the degree distribution in such a way that the latter traverses though one of the critical points.
 Consider a process on the edge-coloured network that thinners the network by randomly removing edges with probability $1-p$, or equivalently, by keeping edges with probability $p$.
 Such removal of edges alters the degree distribution, so that:  $\boldsymbol \mu_0' =p\boldsymbol \mu_0,  \boldsymbol M' = p \boldsymbol M$ and  $(\boldsymbol T_i')_{j,l} = p^2(\boldsymbol T_i)_{j,l} + p (1 - p) M_{j,i} \delta_{j,l}.$ As shown in SI~5 by plugging $\boldsymbol M'$ into the criticality criterion \eqref{eq:criterion} one obtains the following $p$-dependant criterion:
 the edge-coloured network features the critical behaviour at $p=p_\text{c}\in(0,1]$ if there is vector $\mathbf v$ for which, 
\begin{equation}\label{eq:criterionPP}
\mathbf v \in \ker [  p_\text{c}\boldsymbol  M -\boldsymbol I],\;\text{and}\; \frac{\mathbf v}{| \mathbf  v|} \geq 0.
\end{equation}
An alternative way of looking at equation \eqref{eq:criterionPP} is reformulating this criterion as an eigenvalue problem: 
the edge-coloured network features the critical behaviour at $p_\text{c}=\lambda^{-1}$ if the following conditions hold true:\\
1. $(\lambda, \; \mathbf v)$   is an eigenpair of $\boldsymbol  M,$\\
2. $ \lambda >1,$\\
3. $ \mathbf v $ is non-negative when normalised, \emph{i.e.} $\frac{\mathbf v}{|  \mathbf  v|}\geq0$.\\
These three conditions provide a generalisation to a percolation criterion that has been introduced elsewhere   \cite{hackett2016} (equation 7).
Indeed, suppose $\boldsymbol M_{i,j}>0$, then according to the Perron-Frobenius theorem, $\boldsymbol M$ has exactly one positive eigenvector and this eigenvector corresponds to the largest eigenvalue so that the third condition of the above-mentioned list is automatically satisfied. In which case, one can write a simplified criterion:
\begin{equation}\label{eq:criterion_simpl}
p_\text{c}=\lambda^{-1},\;\lambda = \rho(\boldsymbol M)\geq1.
\end{equation}
The essential difference between condition \eqref{eq:criterion_simpl} and the generalisation \eqref{eq:criterionPP} lies in the fact that
equation~\eqref{eq:criterion_simpl} predicts at most one critical point $p_\text{c}$, whereas  equation~\eqref{eq:criterionPP} recovers multiple critical points, if such exist.

\subsection{The hierarchy of critical points and secondary phase transitions}
To this end, our notion of connectivity was based on a multicoloured path, i.e. the path that consists of any colours form index set $\{1,\dots,N\}$. 
By excluding some of the colours, one may define stronger notions of connectivities.
Any non-empty subset  $S\subset\{1,\dots,N\}$  gives rise to a valid definition of a path, and all possibilities collected together comprise a power set $\mathcal P \{1,\dots,N\}\setminus \{\emptyset\}$, \emph{i.e.} the set of all subsets. 
This power set features a natural hierarchy as defined by inclusion relation ``$\subset$''. For example, let $N=3,$ then if two nodes are connected with a path having colours from $\{1,2\},$ this nodes are also connected by a weaker notion of the path that may include colours from $\{1,2,3\},$ since $\{1,2\} \subset \{1,2,3\}$. 
Every connectivity notion $S=\{i_1,i_2,\dots,i_k\}\subset \{ 1,\dots,N\}$ is associated with a distinct notion of criticality, and the inclusion of these subsets imposes a certain ordering of critical points $p_S$, for instance, in the above example: $p_{\{1,2,3\}} \leq p_{\{1,2\}}$.
 Criterion \eqref{eq:criterionPP} can be adjusted to detect these $S$-criticalities:
let $\mathbf x_S$ is the indicator vector for colour subset $S$ and $(\lambda, \mathbf v)$ is an eigenpair of 
\begin{equation}\label{eq:XSM}
\text{diag}\{ \mathbf x_S\} \boldsymbol M \text{diag}\{  \mathbf x_S\},
\end{equation}
then the model features $S$-criticality at $p_S=\lambda^{-1}$ if $\frac{\mathbf v}{|\mathbf v|}\geq0.$
In this way, one has means of identifying all critical points $p_S$ and all colour subsets that correspond to them.   

Although, recovering all $p_S$ requires solving $2^N-1$ eigenvalue problems as defined by equation~\eqref{eq:XSM},
 much can be said on how $p_S$ are distributed when $\boldsymbol M$ is a diagonal-dominant matrix.
In this case, the critical points are grouped into batches, and the locations of this batches are given by the Gershgorin circle theorem: each critical point $p_S$ belongs to at least one of the following intervals:
\begin{equation}
\label{eq:Gershgorin}
\left[ \max(M_{i,i} +r_i,1)^{-1},  
\max(M_{i,i} -r_i,1)^{-1}\right], \; r_i =\sum\limits_{j\neq i} M_{i,j}, \;i=1,\dots,N.
\end{equation}

\subsection{Colour dependant percolation}
Instead of single percolation parameter $p$, consider a vector $\boldsymbol p$ composed of $p_i$, the probabilities that $i$-coloured edge is not removed.
Colour-dependant percolation alters the degree distribution parameters in the following fashion: 
$
\boldsymbol \mu_0' = \text{diag}\{ \mathbf p \}  \boldsymbol \mu_0, 
$
$
\boldsymbol M' = \text{diag}\{ \mathbf p \}  \boldsymbol M 
$,
$\boldsymbol T_i'= \text{diag}\{\mathbf p\}\boldsymbol T_i\text{diag}\{\mathbf p\}+\text{diag}\{\mathbf p\} \text{diag}\{1-\mathbf p\}\text{diag}\{ M_{1,j},\dots,M_{N,j}\}.$
The derivations of this transformations can be found in SI~6.
By plugging $\boldsymbol M'$ into equation~\eqref{eq:criterion} one obtains the criterion for colour-dependent percolation: edge-coloured network features the critical behaviour at $0<\boldsymbol p<1$ if and only if
 \begin{equation}\label{eq:criterionPP2}
\mathbf v \in \ker [\text{diag}\{ \boldsymbol p\}  \boldsymbol M-\boldsymbol I],\;\text{and}\; \frac{\mathbf v}{| \mathbf  v|} \geq 0.
\end{equation}
 The latter criterion can be viewed as a parameter equation for a surface placed in an $N$-dimensional space, and, unlike in the case of simple percolation, one cannot reduce criterion \eqref{eq:criterionPP2} to an eigenvalue problem.

\subsection{Description of the giant component}
The node-size of the giant component $g_\text{node}$ is the probability that a randomly sampled node belongs to the giant component.
This probability can be written in terms of expected values of the degree distribution,
\begin{equation}\label{eq:gnode}
g_\text{node} = 1-\mathbb E[ \mathbf s^\mathbf k ],\; \mathbf s=(s_1,s_2,\dots,s_n)^\top,
\end{equation}
where $s_i $ are obtained by solving
$$ s_i = \frac{\mathbb E[ k_i \mathbf s^{\mathbf k-\mathbf e_i} ]}{\mathbb E[ k_i ]}, i=1,\dots,N,$$
 and $\mathbf e_i$ are the standard basis vectors. 
In a similar fashion, the edge-size of the giant component, the probability that a randomly sampled edge of colour $i$ is a part of the giant component, is given by
$g_i = 1-s_i^2$. The weight-average size of finite connected components $w_\text{avg}=\frac{\mathbb E[n^2]}{\mathbb E[n]}$ is given by:
\begin{equation}
\label{eq:w_avg}
w_\text{avg} = \frac{\mathbf s^\top \boldsymbol D[ \boldsymbol I - \boldsymbol X(\mathbf s) ]^{-1} \mathbf s }{1-g_\text{node}}+1,
\end{equation}
where $\boldsymbol D = \text{diag}\{\mathbb E[ k_i ],\dots,\mathbb E[ k_N ] \}$ and $X(\mathbf s)$ is a matrix function with the following elements
$$X_{i,j}(\mathbf s) = \frac{ 
 \mathbb E [ ( k_i k_j - \delta_{i,j}k_i  ) \mathbf s^{ \mathbf k - \mathbf e_i - \mathbf e_j }  ] 
 }{ \mathbb E [ k_i] },\; i,j=1,\dots,N. $$

When subjected to the bond percolation, $g_\text{node}$ and $w_\text{avg} $ become functions of percolation parameter $p$:
$g_\text{node}(p) = 1-\mathbb E[ (p(\mathbf s_p-1)+1)^\mathbf k ],$
where
\begin{equation}\label{eq:ginat_size}
(s_p)_i = \frac{\mathbb E[ k_i  (p(\mathbf s_p-1)+1)^{\mathbf k-\mathbf e_i} ]}{\mathbb E[ k_i ]}, i=1,\dots,N,
\end{equation}
and
\begin{equation}\label{eq:avg_size}
w_\text{avg}(p) = \frac{\mathbf s_p^\top \boldsymbol D[p^{-1} \boldsymbol I - \boldsymbol X(p(\mathbf s_p-1)+1) ]^{-1} \mathbf s_p }{1-g_\text{node}(p)}+1.
\end{equation}
In this way, the original degree distribution provides enough information to describe how the size of the giant component and the average size of finite components evolve during the percolation progress. Equation \eqref{eq:ginat_size} has been also derived in refs.   \cite{allard2015,hackett2016}, and \eqref{eq:avg_size} generalises similar relation for uncoloured networks   \cite{kryven2018}.
The derivations for equations \eqref{eq:gnode}-\eqref{eq:avg_size} are given in SI~8.

\subsection{Stochastic simulations of percolating coloured networks}
\begin{figure}[htbp]
\begin{center}
\includegraphics[width=\textwidth]{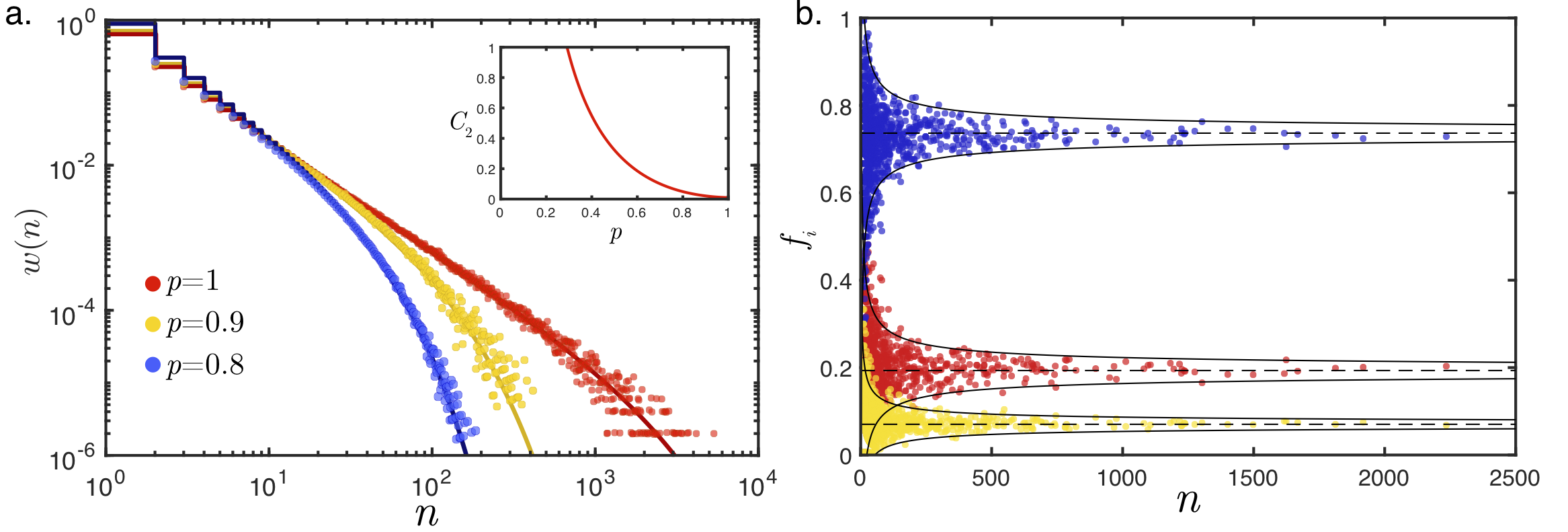}
\end{center}
\caption{ 
{\bf Simple bond percolation in a sub-critical coloured network.}
{\bf a.} Simulated size distributions of connected components (scatter plots) are compared with the theoretical asymptotes (solid lines).
 Inset: dependance of the exponent coefficient $C_2$ on the percolation probability $p$. 
{\bf b.} Fraction of coloured edges $f_i$ that belong  to components of size $n$. Scatter plots: simulated networks (1-red,  2-blue, 3-yellow). Dashed lines: theoretical mean values. Solid lines: theoretical two-sigma standard deviation. }
\label{fig:comb1}
\end{figure}
In this section, we compare the asymptotic theory and stochastic simulations for a few peculiar examples.
Consider a configuration model with three colours that is defined by
\begin{equation}
\label{eq:EX1}
 u( \mathbf k ) = C\, \text{Poiss}\Big( 2k_1 + 3k_2 + 4k_3, 3 \Big),\; \mathbf k \neq0, 
\end{equation}
where $\text{Poiss}\big( k, \lambda \big):= e^{-\lambda}\frac{\lambda^k }{k!}$ is the Poisson mass density function and $C$ ensures the appropriate normalisation.
From equations~\eqref{eq:M},\eqref{eq:T} one obtains matrices $\boldsymbol \mu_0, \boldsymbol M,$ $\boldsymbol T_1,\boldsymbol T_2$ and $\boldsymbol T_3$, which define the asymptotical properties of the network, and from equation~\eqref{eq:simple_asm} one obtains the asymptote of the size distribution:
\begin{equation} \label{eq:ex1:w}
w_\infty(n)=0.7992 n^{-3/2}e^{ -0.00043 n}.
\end{equation}
Figure~\ref{fig:comb1}a compares this asymptote to the size distribution obtained from the simulations. 
This asymptote appears to be almost scale-free. This is in accordance with the close-to-zero value of $C_2=0.00043$: the network is very close to the critical point, however, one cannot assess if the network is just below or just above the phase transition by simply looking at the asymptote. Since Perron-Frobenius conditions are satisfied for $\boldsymbol M$, we can safely apply the simplified test \eqref{eq:criterion_simpl}: the network is below its critical point since $\rho(M) < 1$.
Simple bond percolation with occupancy probabilities $p=0.9$ and $p=0.8$ gives a progressively faster decreasing size distribution.

The colour fractions in the non-giant components settle down on an uneven proportion in components of large size $n\gg1$. 
According to equation~\eqref{eq:conf_dist}, a connected component contains, in average, $19\%$ edges of colour 1, $74\%$ edges of colour 2, and $7\%$ edges of colour 3.
 The covariance matrix of this distribution, as defined in equation~\eqref{eq:Sigma}, features negative correlation and vanishes when $n$ is large: 
\begin{equation} \label{eq:ex1:Sigma}
\boldsymbol \Sigma = \frac{1}{n}
\begin{bmatrix}
          \;\;\;0.22 &  -0.20&   -0.02\\
         -0.20 &   \;\;\,0.25&   -0.05\\
         -0.02 &  -0.05&    \;\;\;0.07
\end{bmatrix}.
\end{equation}
Figure~\ref{fig:comb1}b  compares the theory with simulated colour fractions.
At first glance, this tendency may seem to be contra-intuitive: the larger component is, the less random structure it attains. 
The intuition behind this phenomena lies in the fact that since the size of the component is conditioned to be a large number $n$, the model finds the optimal balance between two contradictory forces: selecting as many nodes of high degree as possible versus selecting nodes that are most abundant according to the degree distribution, which are in this case the nodes with small degree and specific configuration of colours. 
 The variance of the spread present in the scattered data in Fig.~\ref{fig:comb1}b is well explained by the theory. This spread is an outcome of the fact that finite components in infinite systems feature fluctuations in their internal structure.

\begin{figure}[htbp]
\begin{center}
\includegraphics[width=1.05\textwidth]{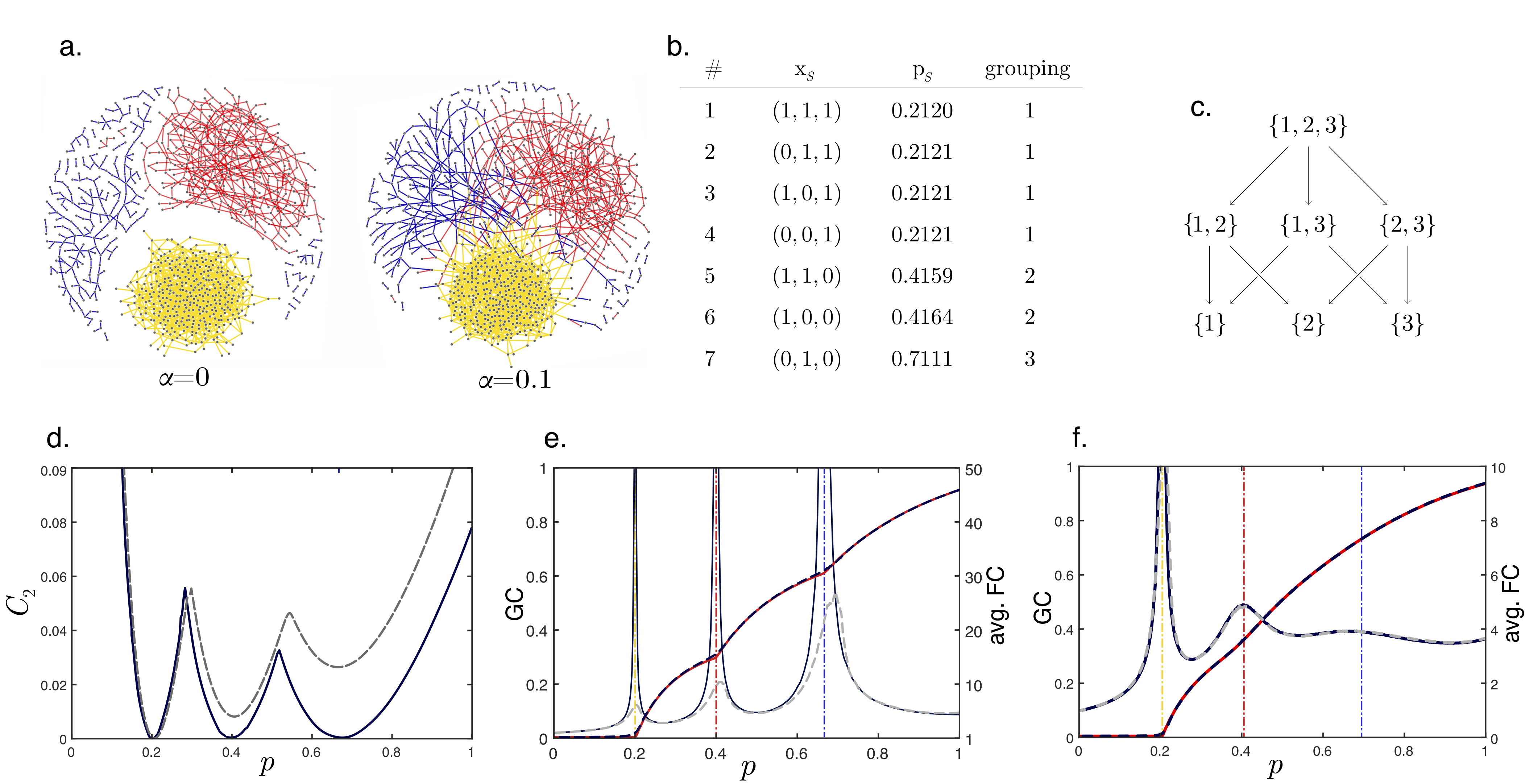}
\caption{
{\bf A coloured network with multiple phase transitions.}
{\bf a,} 
Samples of network topologies corresponding to the degree distribution given by equation \eqref{eq:EX2} with $\alpha=0$ and $\alpha=0.1$.
{\bf b,} The list of all secondary phase transitions and the colour subset indicators that correspond to them. 
{\bf c,} The partial order of all secondary phase transitions in a network with three colours. An arrow points from an earlier-occurring phase transition to a later-occurring one.
{\bf d,} Dependance of the exponent coefficient $C_2$ on the percolation parameter $p$ for two cases of the degree distribution: $\alpha=0$ (solid line), and $\alpha=0.1$ (dashed line). 
{\bf e,f,} Left axis: The total size of all giant components as a function of percolation parameter $p$ as obtained from the theory (red solid line) and simulations (black dashed line). Right axis: The weight-average size of finite connected components as obtained form the theory (black solid line)
and simulations (grey dashed line). These results are obtained for degree distributions with: {\bf e,} $\alpha=0$, and {\bf f,} $\alpha=0.1$.
The vertical guide lines provide theoretical vales for phase transition points.
 }
\label{fig:comb2}
\end{center}
\end{figure}

In the next example, we consider degree distribution
\begin{equation}\label{eq:EX2}
u(\mathbf k) =C\begin{cases}
  \text{Poiss}( k_1, 1.5 ), &  k_2,k_3 = 0;\\
  \text{Poiss}( k_2, 2.5 ), &  k_1,k_3 = 0;\\
  \text{Poiss}( k_3, 5 ),   &  k_1,k_2  =0;\\
  \alpha,                   &  k_1,k_2,k_3 = 1.
  \end{cases}
\end{equation}
When $\alpha=0$, each node has solely links of one colour, so that the whole system is a composition of three unicoloured networks. This fact results in a somewhat exotic situation when multiple giant components can stably coexist, which is an attractive phenomenon from the perspective of many applied disciplines\cite{ben2005b}.
When $\alpha>0$, some nodes have links with many colours and the whole network contains, if any, one giant component. 
Nevertheless, as Fig.~\ref{fig:comb2}a shows, the network is modular in both cases, which points towards the utility of coloured edges for modelling networks with community structures. 

In the case $\alpha=0$, the dependence of $C_2$ on the percolation probability $p$, as given by the solid line in Fig.~\ref{fig:comb2}d, reveals that such criticalities occur three times at $p_\text{c}=\frac{1}{5},\; \frac{2}{5}$, and $\frac{2}{3}$.
Fig.~\ref{fig:comb2}e shows that the weight-average component size is singular at these critical points.
The values of $p_\text{c}$ can be easily deduced from matrix $\boldsymbol M$, which is diagonal when $\alpha=0$:
 $$\boldsymbol M = 
 \begin{bmatrix}
         2.5 &  0   &  0\\
         0   &  1.5 &  0\\
         0   &  0   &  5
\end{bmatrix}.
$$
In this case, the diagonal elements of $\boldsymbol M$ are also its eigenvalues, and all three eigenvectors are positive when normalised. So that, according to the criterion \eqref{eq:criterionPP}, all three eigenvalues are associated with valid phase transitions. Since all off-diagonal elements are zeros, one can invoke equation~\eqref{eq:Gershgorin} to show that all secondary phase transitions coincide with the primary critical points, the complete list is given  in SI~9. Note, that when $\alpha=0$, matrix $\boldsymbol M $ contains zero elements, and therefore the simplified phase transition criterion \eqref{eq:criterion_simpl} is not applicable.

Setting $\alpha=0.1$ in degree distribution \eqref{eq:EX2} perturbs $\boldsymbol M$ so that it becomes  a full matrix with small off-diagonal elements,
$$
\boldsymbol M = 
 \begin{bmatrix}
    2.4    &  0.1 &  0.02\\
    0.04   &  1.4 &  0.02\\
    0.04   &  0.1 &  4.71
\end{bmatrix}.
$$
Although the total size of the giant components, as shown in Fig.~\ref{fig:comb2}e and Fig.~\ref{fig:comb2}f differs only a little,
the eigenvalue decomposition of $\boldsymbol M$ shows that there is only one positive eigenvector, and therefore, one phase transition.
This observation is supported by the fact that $C_2(p)=0$ only once, as shown by the dashed line in Fig.~\ref{fig:comb2}d, and that 
the weight-average component size is singular only at $p\approx 0.21$, as indicated in Fig.~\ref{fig:comb2}f. 
One can yet observe that $C_2(p)$ has two local minima at the locations where the case of $\alpha=0$ features phase transitions.

Since $\boldsymbol M$ is strongly diagonal dominant for $\alpha=0.1$, the secondary critical points appear in groups. The full list, as obtained from criterion \eqref{eq:XSM}, is given in Fig.~\ref{fig:comb2}b. Phase transitions associated with three colours feature the hierarchy as indicated by the partially ordered set in Fig.~\ref{fig:comb2}c, one may also think of Fig.~\ref{fig:comb2}b as a linear sorting of this partial order.

\begin{figure}[htbp]
\begin{center}
\includegraphics[width=0.5\textwidth]{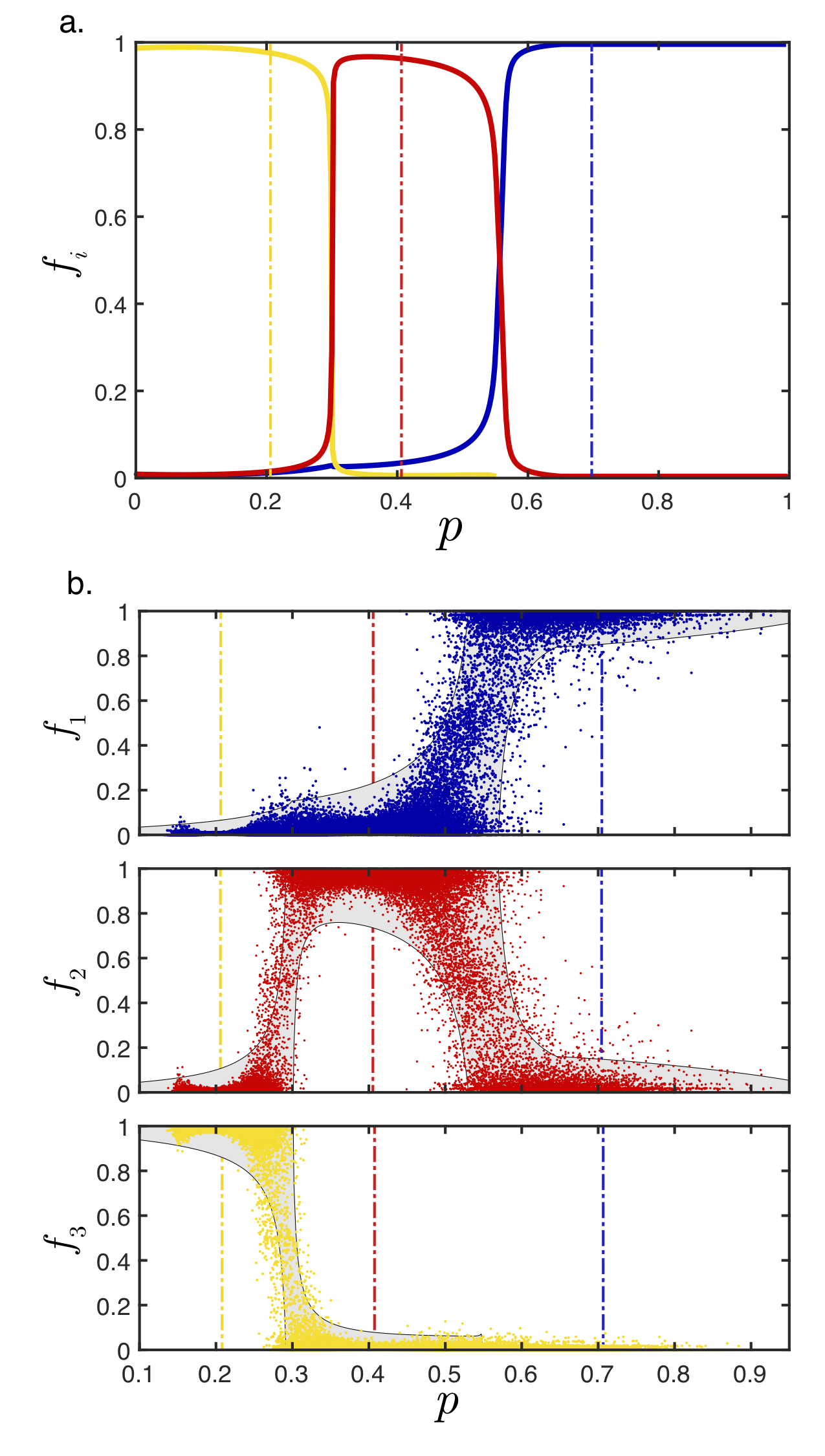}
\caption{
{\bf Colour switching in finite components.}
{\bf a,} Theoretical expected faction of colours in finite components as a function of percolation probability $p$ features a switching behaviour in the coloured network with $\alpha=0.1$.
{\bf b, } Comparison of colour fractions $f_1,f_2,$ and $f_3$ in components of size $n=50$: a generated network (scatter plot) versus theoretical two-sigma standard deviation (shaded areas).
The vertical guide lines provide theoretical vales for phase transition points.
 }
\label{fig:comb3}
\end{center}
\end{figure}

Probably the most surprising implication of the equation describing the internal structure of connected components is not that the colour fractions settle on an uneven ratio, as depicted in Fig.~\ref{fig:comb1}b, but that this ratio peculiarly evolves as a function of $p$. 
For instance, in the case of $\alpha=0.1$, the colour fractions $f_1,f_2,$ and $f_3$ feature a switching behaviour.
Non of the switching points coincides with the phase transitions, but as Fig.~\ref{fig:comb3}a reveals, they are rather equidistant  from the critical points.
 This trend may be exploited to device early warning strategies  that, similar to those devised in  \cite{squartini2013}, detect proximity of a phase transition in empirical networks.
In  contrast to large components, in which the colour fractions converge to precise values as shown in Fig.~\ref{fig:comb1}b, the structure of small components is not deterministic but features fluctuations. The variance of these fluctuations is predicted by the theory and does not vanish in large networks. Figure~\ref{fig:comb3}b compares the simulated data for components of size $n=50$ against the theoretical standard deviation.

\begin{figure}[htbp]
\begin{center}
\includegraphics[width=0.65\textwidth]{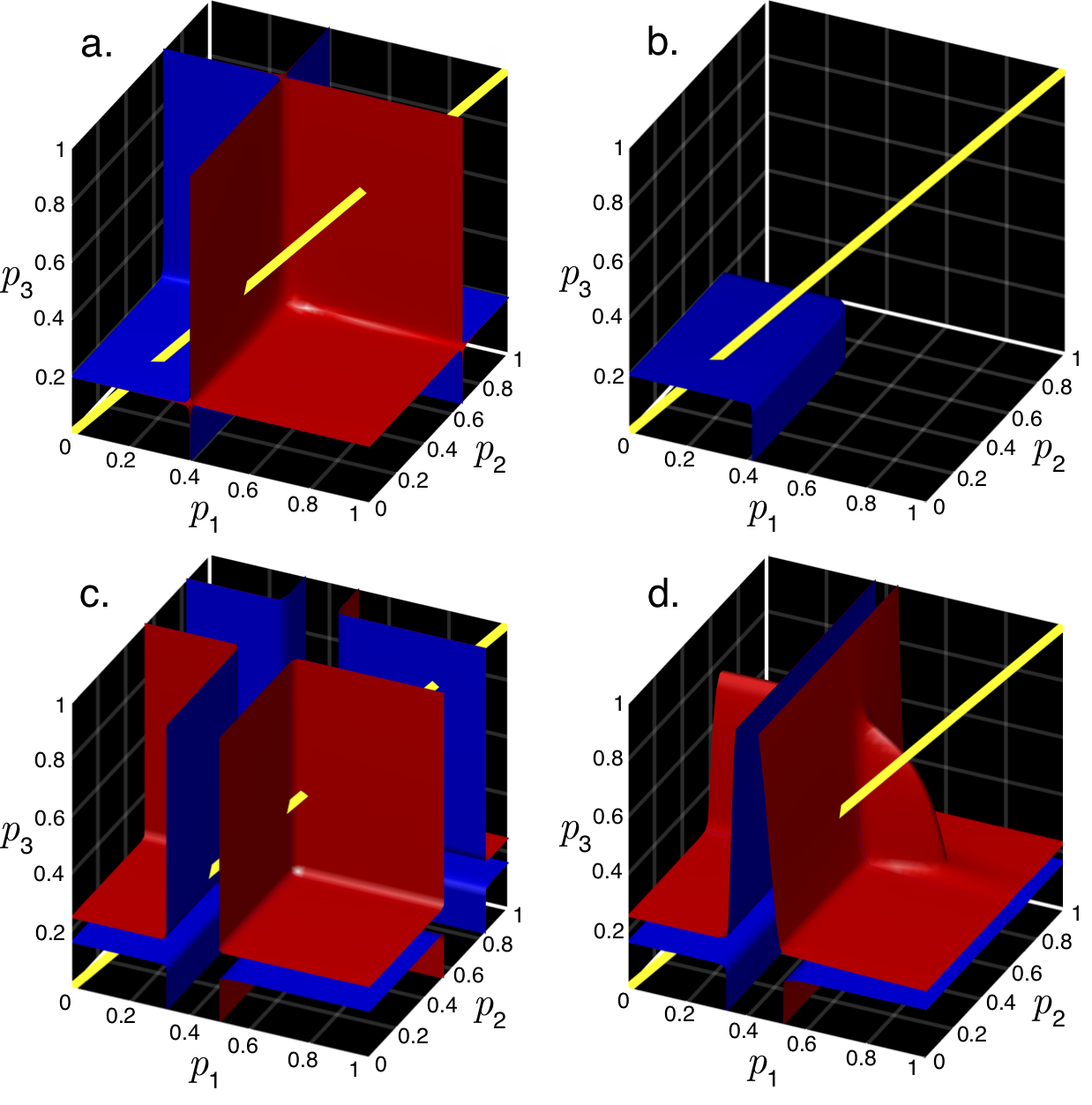}
\caption{
{\bf Colour-dependant percolation in a network with three colours.}
{\bf a,b,} Surfaces containing critical configurations of vector $\mathbf p$ in the network model with: {\bf a,} $\alpha=0$ and {\bf b,} $\alpha=0.1$.
{\bf c,d,} Manifolds containing $\mathbf p$-configurations for which $C_2<0.02$ are computed  for: {\bf c,} $\alpha=0$ and {\bf d.} 
$\alpha=0.1$. 
 The red-blue colouring is used for better visual distinction between the surface's sides in {\bf a,} and {\bf b,} and manifold's interior/exterior in {\bf c,} and {\bf d.} 
 }
\label{fig:comb4}
\end{center}
\end{figure}

 In colour-dependent percolation, one investigates the properties of the percolated network as a function of probability vector $\mathbf p =(p_1,p_2,p_3)$. 
All configurations of $\mathbf p$ amount to the volume of a unit cube. 
  Figure~\ref{fig:comb4} presents the regions of this parameter space where the network becomes critical, that is $C_2=0$, or close-to-critical, $C_2$ is small. This configurations are recovered by numerically solving equation \eqref{eq:criterionPP2} that parametrises the corresponding manifolds.
One can see that what appeared as single curves in Fig.~\ref{fig:comb2}d,e for simple percolation, is now a surface placed in the unit cube, see Fig.~\ref{fig:comb4}.
 When $p_1=p_2=p_3$, which corresponds to a diagonal of this unit cube as indicated by the yellow line, the colour-dependant and simple percolations are equivalent.
Fig.~\ref{fig:comb4}b shows that in the case $\alpha=0.1$, the critical points form a box-shaped surface, whereas $\alpha=0$ changes this surface into a more complex shape, see Fig.~\ref{fig:comb4}a. Note, in Fig.~\ref{fig:comb4}a the yellow line intersects this surface three times, which corresponds to the three phase-transition points that are also observed in Fig.~\ref{fig:comb2}e. 

Although there is a conceptual difference between the cases presented in Fig.~\ref{fig:comb4}a,b, one would expect that the actual networks should be close in some sense. Indeed, Fig.~\ref{fig:comb4}b is obtained by perturbing the network represented in Fig.~\ref{fig:comb4}a with a small parameter $\alpha$.
This similarity can be highlighted if we depart from our strict definition of criticality, $C_2=0$, in favour of a weaker criterion: $C_2<0.02$. All configurations of $\mathbf p$ that satisfy $C_2<0.02$ form  together the three-dimensional manifolds presented in Fig.~\ref{fig:comb4}c,d: one can see that these manifolds do bear resemblance. 
Now, the time evolution the colour-dependant percolation  can be represented with a path $\mathbf p(t),\;t\in[0,1]$ that starts at $\mathbf p(0)=(1,1,1)^\top$, which corresponds to the intact network, and ends at $\mathbf p(1) = (0,0,0)^\top$ --  a completely disintegrated one. Furthermore, the question of how an individual network responds to percolation is about how this path relates to the above-described geometric structures, and it turns out that colour-dependant percolation can lead to completely new behaviours that are not observed in simple percolation.
\begin{figure}[htbp]
\begin{center}
\includegraphics[width=\textwidth]{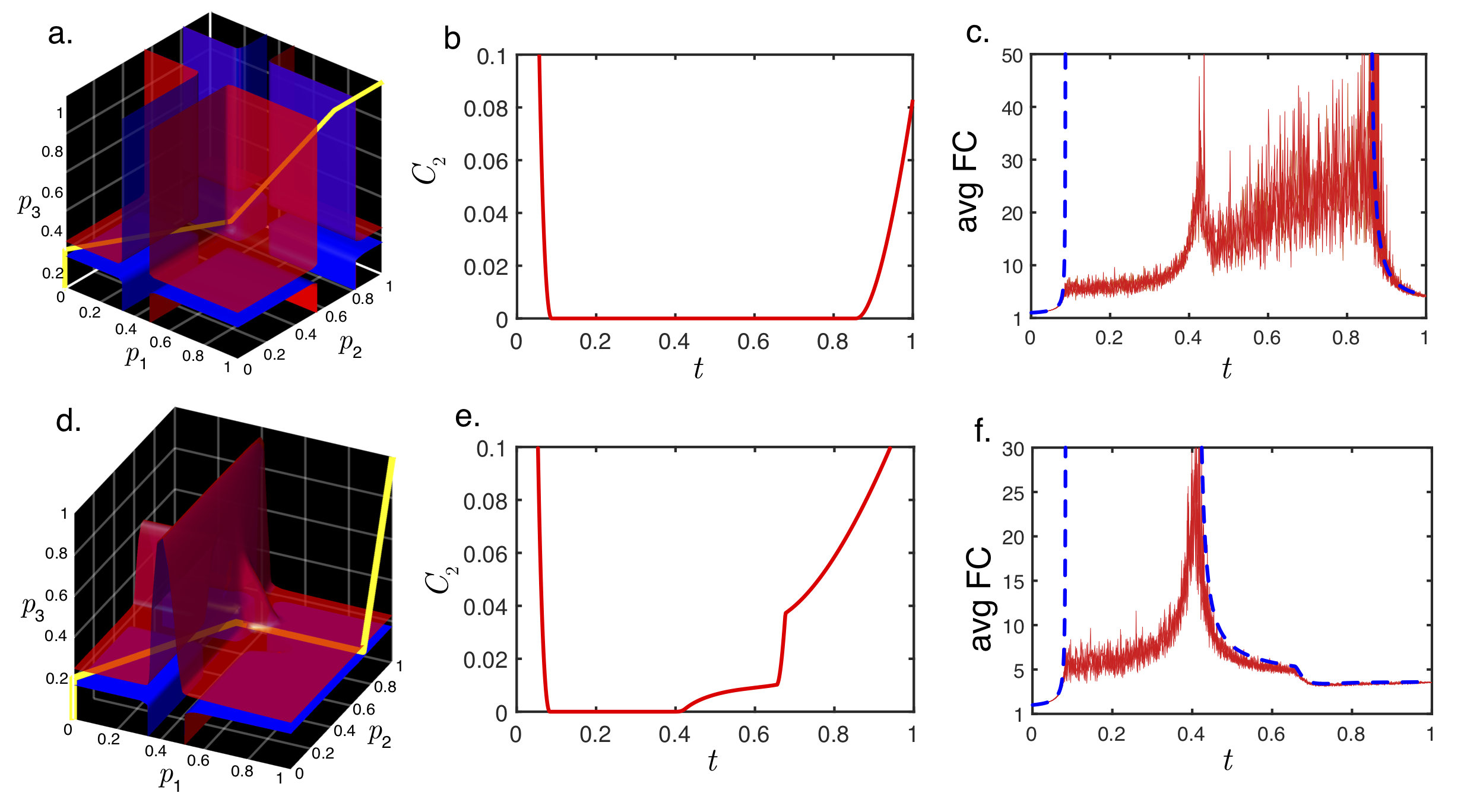}
\caption{
{\bf Colour-dependant percolation that features a wide critical window.}
 Networks corresponding to the degree distribution defined by equation \eqref{eq:EX2} with: {\bf a,b,c,} $\alpha=0$ and  {\bf d,e,f,}  $\alpha=0.1$. 
{\bf a,d,} optimal percolation paths $\boldsymbol p(t)$ together with the reference surfaces on which $C_2=0.02$.
{\bf b,e,} the $C_2$-profiles corresponding to the optimal paths.
{\bf c,f,} comparisons of the theoretical weighted-average component size profiles against the corresponding profiles extracted from stochastically generated networks with $10^5$ nodes.}
\label{fig:comb5}
\end{center}
\end{figure}
In order to demonstrate this fact, we devise a path $\boldsymbol p(t)$ in such a way that $C_2(\boldsymbol p)$ is minimal on it. This is achieved by performing minimisation  $\int\limits_0^1C_2(\boldsymbol p (t))\text{d}t \to \min$. Such optimal paths for the networks with $\alpha=0$ and $\alpha=0.1$ are illustrated in Figs.~\ref{fig:comb5}a,d, whereas Figs~\ref{fig:comb5}b,e depict the corresponding profiles  of $C_2(t)$. One immediately notices that in these examples the profiles of $C_2(t)$ vanish not at discrete points, as was the case with simple percolation, but on intervals of continuum. Everywhere on this intervals the network is critical: that is the sizes of connected components feature the scale-free behaviour and thus the weight-average component size is infinite.
As Fig.~\ref{fig:comb5}c,f shows, this theoretical considerations are also supported by numerically generated networks.  Generated networks comprise of finite number of nodes, and therefore they cannot feature infinite average component size. However, this quantity features macroscopic fluctuations within the critical window and diverges to infinity with growing system size. Remarkably, the width of this critical window does not shrink to zero in infinite systems.  This phenomenon constitutes a new type of phase transitions in coloured networks: the phase transitions with wide critical windows.  Within this critical window the average size of a connected component features macroscopic fluctuations.

\section{Conclusions}
In this paper, edge colours provide an abstraction of an additional layer of information that a network can be equipped with, but also an abstraction of various network structures.
The colours may represent an affiliation to communities, multiplexity, different types of interactions, assortative/disassortative relationships, and other aspects prevalent to complex networks.
We have shown that colour dependencies may amplify failures in connectivity or make the collapse of the network less sudden.
 This is why this theory provides the foundation for answering the key question of complex networks science: ``How we can we economically design robust multilayer networks of infrastructures or financial networks with trustworthy links?'' \cite{bianconi2014b}.
In many ways a modeller exploit the above-described geometric interpretation of colour-dependant percolation in sake of network design and control. 
By following similar motivations to the ones in Ref. \cite{osat2017}, one might aim to optimise the percolation path so that minimal/maximal number of edges is removed before percolation reaches the phase transition, or for a fixed path, one might optimise the network robustness so that the phase transition manifold avoids the path as much as possible. Another objective, reducing the sharpness of the phase transition\cite{son2011,buldyrev2010}, can be achieved by reducing the angle between the path and the surface at the intersection point. In fact, as we have demonstrated by an example, one may even construct a path that do not immediately intersect the manifold but stays inside for a long time, and therefore, keeps the network in the critical window. 
The latter observation shifts the paradigm of the critical window itself, as it demonstrates that even in infinite systems, the critical window may feature a non-vanishing width.

\section{Methods}
\subsection*{Expectations of the multi-degree distribution}
In a network with $N$ colours, the degree distribution $u(\boldsymbol k),\;  \boldsymbol k = (k_1,\dots,k_N)$ is the probability that a randomly selected node bears $k_i$ edges of colour $i$. The expected value of $f(\boldsymbol k)$ given degree distribution $u(\boldsymbol k),$ is defined by the following sum: $$\mathbb E[f(\boldsymbol k)]: = \sum\limits_{\boldsymbol k\geq0} f(\boldsymbol k) u(\boldsymbol k).$$  
It is convenient to gather all expected values for $k_i$ into the expected vector, $\boldsymbol \mu_0 =(\mathbb E[k_1],\mathbb E[k_2],\dots,\mathbb E[k_N])^\top$.
If instead of selecting a node at random, one selects a node at the end of a randomly chosen, $i$-coloured edge, the corresponding degree distribution is called $i$-biased and  given by $(\mathbf k + \mathbf e_i)\frac{u(\mathbf k + \mathbf e_i)}{\mathbb E [k_i]}$.
The expected vectors of the $i$-biased degree distributions are given by the columns of the following $N\times N$ matrix,
\begin{equation}\label{eq:M'}
M_{i,j}  = \frac{ \mathbb E[ k_i k_j ]}{\mathbb E[ k_j]}-\delta_{i,j},
\end{equation}
and the covariance matrices of $i$-biased degree distributions are given by
\begin{equation}\label{eq:T'}
(\boldsymbol T_i)_{j,l} = \frac{\mathbb E[ k_i k_j  k_l ]}{\mathbb E[ k_i]}
- \frac{\mathbb E[k_i k_j ]\mathbb E[k_i k_l ]}{\mathbb E[ k_i]^2},\; j,l=1,\dots,N.
\end{equation}

\subsection*{Coefficients of the asymptote}

Consider an axillary function 
\begin{equation}\label{eq:f}
f(\mathbf z) :=  \frac{1}{2} [1, \mathbf z^\top] \boldsymbol K^\top   \boldsymbol \sigma_{\mathbf z} ^{-1} \boldsymbol K 
\begin{bmatrix}
1\\
\mathbf z
\end{bmatrix},
\end{equation}
where $\mathbf z$ is a vector of dimension $N-1$, $\mathbf K= (\boldsymbol M-\boldsymbol I )\mathbf A^{-1},$ 
$$\sigma_{\mathbf z} = (1-|\mathbf z|)\boldsymbol T_1 + \sum\limits_{i=1}^{N-1} z_i \boldsymbol T_{i+1},\; | \mathbf z | := \sum\limits_{i=1}^{N-1} z_i, $$
and
$$A_{ i, j }  =
\begin{cases} 1 & i=1,\\
 \delta_{ i, j }  & i>1.
\end{cases} 
$$
Let $\mathbf z^* = \arg \min f(\mathbf z),\;  \mathbf z^*>0,\; |\mathbf z^* | \leq 1,$
is the point where $f(\mathbf z)$ reaches its minimum value, then the size distribution \eqref{eqA1:multiplex} features the following asymptote at the tail: 
\begin{equation}
\label{eq:simple_asm_M1}
w_{\infty}(n)= C_1 n^{-\sfrac{3}{2} } e^{- C_2 n},
\end{equation}
where the coefficients are given by
 $$C_1 = \det(\boldsymbol M - \boldsymbol I)  \text{tr}(\text{adj}(\boldsymbol Q) \boldsymbol R )\left( \frac{ \det[  \mathbf H^{-1}_{\mathbf z^*} ]}{ 2 \pi  \det[ \boldsymbol\sigma_{\mathbf z^*}]} \right)^{1/2}\!\!
e^{-[1, \mathbf z^{*\top}] \boldsymbol K^\top   \boldsymbol \sigma_{\mathbf z^*} ^{-1}  \boldsymbol \mu_0}$$
and
$$C_2 = \frac{1}{2} [1, \mathbf z^{*\top}] \boldsymbol K^\top   \boldsymbol \sigma_{\mathbf z^*} ^{-1} \boldsymbol K 
\begin{bmatrix}
1\\
\mathbf z^*
\end{bmatrix}.$$
Here  $Q_{i,j} =\delta_{i,j}-[1,\mathbf z^*]^\top \boldsymbol K^\top \sigma^{-1}_{\mathbf z^*}\text{adj}(\boldsymbol M - \boldsymbol I)\boldsymbol  T_i\,  \boldsymbol e_j,$ and
$R_{i,j} =  \boldsymbol \mu_0^\top \, \sigma^{-1}_{\mathbf z^*}\text{adj}(\boldsymbol M - \boldsymbol I)\boldsymbol  T_i\,  \boldsymbol e_j$ are square matrices of size $N\times N$,
and
\begin{multline}
(\boldsymbol H_{\mathbf z})_{i,j}= [1, \mathbf z^\top]  \boldsymbol K^\top    \sigma_{\mathbf z} ^{-1}(\boldsymbol T_{j+1}-\boldsymbol T_1)\boldsymbol \sigma_{\mathbf z} ^{-1}(\boldsymbol T_{i+1}-\boldsymbol T_1)\boldsymbol \sigma_{\mathbf z} ^{-1} \boldsymbol K 
\begin{bmatrix}
1\\
\mathbf z
\end{bmatrix}
+  \boldsymbol e_{i+1}^\top \boldsymbol K^\top \boldsymbol \sigma_{\mathbf z} ^{-1} \boldsymbol K \boldsymbol e_{j+1} \\ 
-  [1, \mathbf z^\top] \boldsymbol K^\top \boldsymbol \sigma_{\mathbf z} ^{-1}(\boldsymbol T_{j+1}-\boldsymbol T_1)\boldsymbol \sigma_{\mathbf z} ^{-1} \boldsymbol K \boldsymbol e_{i+1} 
-  [1, \mathbf z^\top] \boldsymbol K^\top   \boldsymbol \sigma_{\mathbf z} ^{-1} (\boldsymbol T_{i+1}-\boldsymbol T_1)\boldsymbol \sigma_{\mathbf z} ^{-1} \boldsymbol K \boldsymbol e_{j+1}.
\end{multline}
is the Hessian matrix of $f(\mathbf z)$ having size $N-1$ by $N-1$. The standard basis vectors are denoted by $(\boldsymbol e_i)_j = \delta_{i,j}$.
The derivation of these expressions is given in SI~3.

\subsection*{Colour-fraction configuration of a connected component of size $n$ }
Since the total number of edges is $|\mathbf v|=n-1,$ one writes $v_1 = 1-\sum\limits_{i=2}^N v_i,$ and the configurations for the rest of edge types $v_2,v_3,\dots v_N$ obey the following law of large numbers: 
 \begin{multline}\label{eq:conf_dist}
\mathbb P[ \lfloor n z_1\rfloor \leq v_2 \leq\lceil n z_1\rceil  \wedge ,\dots, \wedge 
\lfloor n z_{N-1}\rfloor  \leq v_N \leq\lceil n z_{N-1}\rceil 
 ]  
=
\frac{
e^{-\frac{1}{2} (\mathbf z - \mathbf z^*)^\top [\frac{1}{n} \mathbf H^{-1}_{\mathbf z^*} ]^{-1} (\mathbf z - \mathbf z^*)}
}{\det[2 \pi  \frac{1}{n}\mathbf H^{-1}_{\mathbf z^*} ]^{1/2}}.\end{multline}
Let  $f_i=\frac{v_i}{n-1}$ denote colour fractions in components of size $n$.
The configurations of the colour-fraction vector, $(f_1,f_2,\dots, f_N)$, are normally distributed with mean 
$\mathbf m=  \mathbf z^*$ 
and variance matrix
\begin{equation}\label{eq:Sigma}
\boldsymbol \Sigma = \frac{1}{n} 
\begin{bmatrix}
a& \boldsymbol b^\top\\
\boldsymbol b& \mathbf H_{\mathbf z^*} 
\end{bmatrix},
\end{equation}
where  $a=\sum\limits_{i,j=1}^{N-1} (\mathbf H_{\mathbf z^*})_{i,j}, $ and $\boldsymbol b$ is a column vector of length $N-1$,
$b_i = -\sum\limits_{j=1}^{N-1} (\mathbf H_{\mathbf z^*})_{i,j}.$ The derivation of equations \eqref{eq:conf_dist}-\eqref{eq:Sigma} is given in SI~7.

\begin{acknowledgments}
This work is part of the project number 639.071.511, which is financed by the Netherlands Organisation for Scientific Research (NWO) VENI.
\end{acknowledgments}

\newpage
\part*{Supporting Information}

\title{Supplementary Information for: How does bond percolation happen in coloured networks?}

\author{Ivan Kryven} 
\email{i.kryven@uva.nl}
\affiliation{Van 't Hoff Institute for Molecular Sciences, University of Amsterdam, Science Park 904, 1098 XH Amsterdam, The Netherlands} 
\maketitle

\renewcommand{\top}{\oldtop\!}
\setcounter{equation}{0}
\setcounter{figure}{0}
\setcounter{table}{0}
\setcounter{page}{1}
\makeatletter
\renewcommand{\theequation}{S\arabic{equation}}
\renewcommand{\thefigure}{S\arabic{figure}}
\renewcommand{\thesection}{SI\arabic{section}}
\renewcommand{\thetable}{SI\arabic{table}}

\section{Summary of the asymptotic theory for unicoloured networks } \label{SI:summary}
When there is only one type of edges, $N=1$,  $u(k)$ is simply the probability that a randomly selected node bears $k$ edges. In the configuration model, the size distribution of connected components is given by convolution powers of the degree distribution \cite{kryven2017a}:
\begin{equation}
\label{eqsi:Lagrange1d}
w(n)=\begin{cases}
\frac{\mathbb E [k]}{n-1} u_1^{*n}(n-2),& n>1, \\
u(0) & n=1.
\end{cases}
\end{equation}
Here $u_1(k)=\frac{(k+1)u(k+1)}{\mathbb E [k]}$ is the biased degree distribution\cite{newman2010}, 
the convolution product $d(n)=f(n)*g(n)$ is defined as
$$
d(n):=\sum\limits_{j+k=n}f(j)g(k),\;j,k\geq0,
$$
and the convolution power is defined by induction: $f(k)^{*n}=f(k)^{*n-1} * f(k),\; f(k)^{*0}:=\delta(k)$.
When the degree distribution has a light tail, the equation \ref{eqsi:Lagrange1d} features the universal asymptote \cite{kryven2017a},
\begin{equation}
\label{eqsi:1dasymptote}
w_\infty(n) = C_1e^{-C_2 n} n^{-3/2}.
\end{equation}
The expressions for $C_1$ and $C_2$ are given in terms of the first three moments: $\mathbb E [k]$, $\mathbb E [k^2]$, and $\mathbb E [k^3]$.
Namely,
$C_1 = \frac{\mathbb E [k]^2}{ \sqrt{2 \pi (\mathbb E [k] \mathbb E [k^3]-\mathbb E [k^2]^2) }},$
$C_2 = \frac{(\mathbb E [k^2]-2 \mathbb E [k])^2}{2( \mathbb E [k] \mathbb E [k^3]- \mathbb E [k^2]^2)}$.
At the critical point, when $C_2=0,$ the asymptote \eqref{eqsi:1dasymptote} indicates scale-free behaviour, and features exponent $-\frac{3}{2}$.
The condition $C_2=0$ is equivalent to Molloy and Reed giant component criterion:
$$\mathbb E [k^2]-2 \mathbb E [k]=0.$$

The next section derives a generalisation of this theory for the case of arbitrary number of colours, in which case the coefficients of the N-colour asymptote are derived in terms of mixed moments up to the third order.

\section{Size distribution of connected components with $N$ colours }
In a network with $N$ colours, the degree distribution $u(\boldsymbol k),\;  \boldsymbol k = (k_1,\dots,k_N)$ is the probability that a randomly selected node bears $k_i$ edges of colour $i=1,2,\dots,N.$
Let $\boldsymbol \mu_0$ is the vector-valued expectation of this distribution, 
\begin{equation}\label{eqsi:mu0}
\boldsymbol \mu_0 =(\mathbb E[k_1],\mathbb E[k_2],\dots,\mathbb E[k_N])^\top.
\end{equation}
If one choses a node at the end of a random, $i$-coloured edge instead of choosing a node at random,
 the corresponding degree distribution is called $i$-biased: 
\begin{equation}
u_i(\mathbf k)=(\mathbf k + \mathbf e_i)\frac{u(\mathbf k + \mathbf e_i)}{\mathbb E [k_i]},
\end{equation}
where $\mathbf e_i$ are the standard basis vectors.
The expected column-vectors $\boldsymbol \mu_{i}$ of $i$-biased degree distributions can be expressed in terms of expectations of $u(\mathbf k)$:
\begin{equation}\label{eqsi:mui}
 (\boldsymbol \mu_{i})_j = \frac{ \mathbb E[ k_i k_j ]}{\mathbb E[ k_i]}-\delta_{i,j},\; i,j=1,\dots,N.
\end{equation}
For convenience of notation, we introduce  $\boldsymbol M=(\boldsymbol \mu_{1},\boldsymbol \mu_{2}, \dots,\boldsymbol \mu_{N})
$, a matrix that contains $\boldsymbol \mu_i$ as its columns.
The covariance matrices of $i$-biased degree distributions are given by
\begin{equation}\label{eqsi:T}
(\boldsymbol T_i)_{j,l} = \frac{\mathbb E[ k_i k_j  k_l ]}{\mathbb E[ k_i]}
- \frac{\mathbb E[k_i k_j ]\mathbb E[k_i k_l ]}{\mathbb E[ k_i]^2},\; i,j,l=1,\dots,N.
\end{equation}

In a similar fashion to how Eq. \eqref{eqsi:Lagrange1d} was derived, Ref. \cite{kryven2017b} applies Joyal's theory of species to 
write the exact expression for the  size distribution of weakly connected components: 
\begin{equation}\label{eqA1:multiplex}
w(n)  = \sum_{\substack{k_1+\dots+k_N=n+1\\k_i \geq 0}} \left(\text{det}_*(\boldsymbol D)* u *  u_1^{*k_1}*\dots * u_N^{*k_N} \right)(\mathbf k),
\end{equation}
where the $N$-dimensional convolution $d(\mathbf n)=f(\mathbf n)*g(\mathbf n ),$ is defined as 
\begin{equation}\label{eqsi:convnd}d(\mathbf n): = \sum\limits_{\mathbf j + \mathbf k=\mathbf n}f(\mathbf j)g(\mathbf k),
\; 
\mathbf j, \mathbf k \geq0.
\end{equation}
The sum in Eq.~\eqref{eqsi:convnd} runs over all partitions of the $N$-dimensional vector $\mathbf n$ into two non-negative terms, $\mathbf j$ and $\mathbf k$,  $\det_*(\boldsymbol D)$ refers to the determinant computed with the multiplication replaced by the convolution, and matrix $\boldsymbol D$ has the following elements
\begin{equation}\label{eqsi:D*}
D_{i,j}=\delta(\mathbf k)\delta_{i,j} - [k_j u_i( \mathbf k ) ]* u_i( \mathbf k )^{*(-1)},\; i,j=1,\dots,N.
\end{equation}

\section{Derivations for the size distribution asymptote}
This section derives the asymptote for the equation \eqref{eqA1:multiplex}.
Note that in Eq.~\eqref{eqA1:multiplex}, the arguments of the convolved functions and the convolution powers are both taken from the same vector $\boldsymbol k.$  In order to circumvent this issue, let us  decouple the argument from the powers.
We thus define $G(\boldsymbol k',\boldsymbol k):=u_1(\mathbf k')^{*k_1}*\dots * u_N(\mathbf k')^{*k_N}$. 
To introduce the idea of the asymptotic analysis, we first start with a simplified version of \eqref{eqA1:multiplex}.
\begin{equation}\label{eqA1:multiplex*}
w'(n+1)  = \sum_{\substack{k_1+\dots+k_N=n\\k_i \geq 0}} u_1(\mathbf k')^{*k_1}*\dots * u_N(\mathbf k')^{*k_N} = \sum_{\substack{k_1+\dots+k_N=n\\k_i \geq 0}} G(\boldsymbol k,\boldsymbol k).
\end{equation}
\subsection{Asymptote for the product of convolution powers}

Let $\phi( \boldsymbol \omega )$ denotes the characteristic function of $u( \mathbf k), $
$$\phi( \boldsymbol \omega ) = \sum\limits_{ \mathbf k \geq 0 } e^{\iu \boldsymbol \omega^\top \mathbf k  } u( \mathbf k ),\; \boldsymbol\omega \in \mathbb R^N,\;\iu^2=-1,$$ 
and $\phi_i( \boldsymbol \omega )$ denote the characteristic functions of $u_i( \mathbf k) $.  
 Then, the characteristic function of $G(\boldsymbol k,\boldsymbol k')$ is given by the  product 
\begin{equation}\label{eqsi:psi}
 \psi(\boldsymbol \omega,\mathbf k ) =\phi( \boldsymbol \omega )\prod\limits_{i=1}^N \phi_i^{k_i}( \boldsymbol \omega ),
 \end{equation}
however, as follows from the central limit theorem, $\phi_i^{k_i}( \boldsymbol \omega )$ approaches a limit function when $k_i$ is large:  
$$
\lim_{k_i\to\infty}| \phi_i^{k_i}( \boldsymbol \omega)- \phi_{i,\infty}^{k_i}( \boldsymbol \omega) |=0, 
$$
where
\begin{equation}\label{eqsi:phi_inf}
\phi_{i,\infty}^{k_i}( \boldsymbol \omega) =e^{\iu\, k_i \boldsymbol \mu_i^\top\boldsymbol\omega-\frac{1}{2}  \boldsymbol\omega^\top k_i\boldsymbol  T_i\,\boldsymbol\omega},
\end{equation}
and $ \boldsymbol \mu_i$ and $ \boldsymbol T_i$ are the expected values and covariance matrices as defined by \eqref{eqsi:mui} and \eqref{eqsi:T}.
Applying similar procedure to the equation \eqref{eqsi:psi}, we obtain:
$$
\lim_{k_1,\dots,k_N\to \infty}| \psi(\boldsymbol \omega,\mathbf k ) -\psi_\infty(\boldsymbol \omega,\mathbf k )|=0 ,
$$
where
\begin{equation}\label{eqsi:limCLT}\psi_\infty(\boldsymbol \omega,\mathbf k ) = e^{\iu \boldsymbol\omega (\boldsymbol M \boldsymbol k+ \boldsymbol \mu_0) -\frac{1}{2} \boldsymbol\omega^\top \boldsymbol\sigma(k) \,\boldsymbol\omega},
\end{equation}
and $\sigma(\boldsymbol k)=\sum\limits_{i=1}^N k_i \boldsymbol T_i + o(\mathbf k).$
Inverse flourier transform of Eq.~\eqref{eqsi:limCLT} is easy to obtain since $\psi_\infty(\boldsymbol \omega,\mathbf k )$ itself is the characteristic function of the multivariate Gaussian function:
\begin{equation}\label{eqA:a_inf2}
\begin{aligned}
G(\mathbf k',\mathbf k ) = 
&\det[2 \pi  \boldsymbol\sigma(\mathbf k)]^{-1/2}
e^{ -\frac{1}{2} (\mathbf k'-\boldsymbol M\mathbf k-\boldsymbol\mu_0)^\top \boldsymbol\sigma^{-1}(\mathbf k)( \mathbf k'-\boldsymbol M\mathbf k-\boldsymbol\mu_0) } =\\
&
\det[2 \pi  \boldsymbol\sigma(\mathbf k)]^{-1/2}
e^{ -(\boldsymbol M\mathbf k-\mathbf k')^\top \boldsymbol\sigma^{-1}(\mathbf k) \boldsymbol\mu_0
 -\frac{1}{2}\boldsymbol \mu_0^\top \boldsymbol\sigma^{-1}(\mathbf k) \boldsymbol\mu_0}
e^{ -\frac{1}{2} (\mathbf k'-\boldsymbol M\mathbf k)^\top \boldsymbol\sigma^{-1}(\mathbf k)( \mathbf k'-\boldsymbol M\mathbf k) }. \; 
\end{aligned}
\end{equation}
Consider the following matrix
$$(\boldsymbol A)_{ i, j }  =
\begin{cases} 1 & i=1,\\
 \delta_{ i, j }  & i>1,
\end{cases} 
 $$
and the transformation it induces $\mathbf z' = \frac{1}{n} \boldsymbol A \mathbf k$.
Since $k_1+k_2+\dots+k_N=n$, we have $z'_1=1$ and therefore we may write  $\mathbf z' = \begin{bmatrix}
1\\
\mathbf z
\end{bmatrix}$, where $\mathbf z$ is a vector of dimension $N-1$.
Let us now set $\mathbf k'=\mathbf k$ and introduce the transformation of variables $\mathbf k = n \boldsymbol A^{-1}\begin{bmatrix}
1\\
\mathbf z
\end{bmatrix},$ which replaces the functions appearing in the exponent of \eqref{eqA:a_inf2} with
$$
\frac{1}{2} (\mathbf k'-\boldsymbol M\mathbf k)^\top \boldsymbol\sigma^{-1}(\mathbf k)( \mathbf k'-\boldsymbol M\mathbf k)
 =\frac{1}{2}n [1, \mathbf z^\top] \boldsymbol K^\top   \boldsymbol \sigma_{\mathbf z} ^{-1} \boldsymbol K 
\begin{bmatrix}
1\\
\mathbf z
\end{bmatrix}=nf(\mathbf z),
$$
and
$$
\begin{aligned}
(\boldsymbol M\mathbf k-\mathbf k')^\top \boldsymbol\sigma^{-1}(\mathbf k)\boldsymbol\mu_0+\frac{1}{2}\boldsymbol\mu_0^\top \boldsymbol\sigma^{-1}(\mathbf k)\boldsymbol\mu_0
 =&
  [1, \mathbf z^\top] \boldsymbol K^\top   \boldsymbol \sigma_{\mathbf z} ^{-1} \boldsymbol \mu_0+
 \frac{1}{2n} \boldsymbol \mu_0^\top   \boldsymbol \sigma_{\mathbf z} ^{-1} 
\boldsymbol \mu_0 =\\
 &g(\mathbf z)+ O(n^{-1}),\;g(\mathbf z)= [1, \mathbf z^\top] \boldsymbol K^\top   \boldsymbol \sigma_{\mathbf z} ^{-1}  \boldsymbol\mu_0,
\end{aligned}
$$
where  $\mathbf K= (\boldsymbol M-\boldsymbol I )\mathbf A^{-1}, $ 
and
\begin{equation}\label{eqsi:sigma}
\sigma_{\mathbf z} = (1-|\mathbf z|)\boldsymbol T_1 + \sum\limits_{i=1}^{N-1} z_i \boldsymbol T_{i+1},\; | \mathbf z | := \sum\limits_{i=1}^{N-1} z_i, 
\end{equation}
are independent of $n$.
The Taylor expansion of $f(\mathbf z)$ around the point where this function reaches its minimum value,
\begin{equation}\label{eqsi:z*}
\mathbf z^* = \arg\min  f(\mathbf z),
\end{equation}
 gives:
   \begin{equation}\label{eqA:taylor}
   f(\mathbf z)  = \frac{1}{0!} f(\mathbf z^*) + \frac{1}{1!}\nabla  f(\mathbf z^*)^\top (\mathbf z - \mathbf z^*) +\frac{1}{2!}(\mathbf z - \mathbf z^*)^\top \mathbf H_{\mathbf z^*}  (\mathbf z - \mathbf z^*) + \mathbf R(\mathbf z, \mathbf z^*),
    \end{equation}
  where the $N-1$-by-$N-1$ matrix $ \mathbf H_{\mathbf z^*}$ is the Hessian of $ f(\mathbf z)$  at $\mathbf z=\mathbf z^*$:     
\begin{multline}
(\boldsymbol H_{\mathbf z})_{i,j}= [1, \mathbf z^\top]  \boldsymbol K^\top    \sigma_{\mathbf z} ^{-1}(\boldsymbol T_{j+1}-\boldsymbol T_1)\boldsymbol \sigma_{\mathbf z} ^{-1}(\boldsymbol T_{i+1}-\boldsymbol T_1)\boldsymbol \sigma_{\mathbf z} ^{-1} \boldsymbol K 
\begin{bmatrix}
1\\
\mathbf z
\end{bmatrix}
+  \boldsymbol e_{i+1}^\top \boldsymbol K^\top \boldsymbol \sigma_{\mathbf z} ^{-1} \boldsymbol K \boldsymbol e_{j+1} \\ 
-  [1, \mathbf z^\top] \boldsymbol K^\top \boldsymbol \sigma_{\mathbf z} ^{-1}(\boldsymbol T_{j+1}-\boldsymbol T_1)\boldsymbol \sigma_{\mathbf z} ^{-1} \boldsymbol K \boldsymbol e_{i+1} 
-  [1, \mathbf z^\top] \boldsymbol K^\top   \boldsymbol \sigma_{\mathbf z} ^{-1} (\boldsymbol T_{i+1}-\boldsymbol T_1)\boldsymbol \sigma_{\mathbf z} ^{-1} \boldsymbol K \boldsymbol e_{j+1},
\end{multline} 
  and   $\mathbf R(\mathbf z, \mathbf z^*)$ is the expansion residue.
  Note that the gradient $\nabla  f(\mathbf z^*)=0$ since $\mathbf z^*$ is a minimum of $ f(\mathbf z)$,  and by plugging these expansion into \eqref{eqA:a_inf2} we obtain:
 $$
\begin{aligned}
F_n( \mathbf z)  :=G(\mathbf k,\mathbf k) =&
\det[2 \pi  \boldsymbol\sigma_{\mathbf z}]^{-1/2}
e^{-g(\mathbf z)}e^{  - n   f(\mathbf z)}  =\\
&\det[2 \pi  \boldsymbol\sigma_{\mathbf z}]^{-1/2}e^{-g( \mathbf z)}
e^{-n f(\mathbf z^*)  -\frac{1}{2} (\mathbf z - \mathbf z^*)^\top n \mathbf  H_{\mathbf z^*}  (\mathbf z - \mathbf z^*) - n\mathbf R(\mathbf z, \mathbf z^*).}
\end{aligned}
$$
Now, by multiplying $F_n(\mathbf z)$ with $ 1= \frac{\det[2 \pi  \frac{1}{n} \mathbf H^{-1}_{\mathbf z^*} ]^{1/2}}{\det[2 \pi  \frac{1}{n}\mathbf H^{-1}_{\mathbf z^*} ]^{1/2}}$ and rewriting the sum of exponents  as a product of exponential functions isolates a multivariate Gaussian function in the expression of $F_n(\mathbf z)$:
\begin{equation}\label{eqA:Fn}
F_n( \mathbf z)  = \frac{
e^{-\frac{1}{2} (\mathbf z - \mathbf z^*)^\top [\frac{1}{n} \mathbf H^{-1}_{\mathbf z^*} ]^{-1} (\mathbf z - \mathbf z^*)}
}{\det[2 \pi  \frac{1}{n}\mathbf H^{-1}_{\mathbf z^*} ]^{1/2}}\,
\frac{\det[2 \pi  \frac{1}{n}\mathbf H^{-1}_{\mathbf z^*} ]^{1/2}}{\det[2 \pi  \boldsymbol\sigma_{\mathbf z}]^{1/2}}
\,e^{  -g(\mathbf z)}e^{-\frac{1}{2}n f(\mathbf z^*)  -\frac{1}{2}n \mathbf R(\mathbf z, \mathbf z^*).}
 \end{equation}
Let us now turn back to the  asymptotic analysis of the equation \eqref{eqA1:multiplex*},  which becomes written as
\begin{equation}\label{eqsi:temp}
w(n)'=\sum_{\substack{k_1+\dots+k_N=n-1\\k_i \geq 0}} F_n( \frac{1}{n} \boldsymbol A \mathbf k)=\sum_{\mathbf z\in \Omega_n} F_n( \mathbf z).
\end{equation}
The latter summation is performed over 
 $$\Omega_n:=\{(1,z_1,\dots,z_{N-1})^\top : |\mathbf z| \leq 1, \; z_i =\frac{m}{n},\; m=0,\dots,n\},$$
which, as $n\to \infty$, becomes dense in 
\begin{equation}\label{eqsi:Omegainfty}
\Omega_{\infty}:=\{(1,z_1,\dots,z_{N-1}) : |\mathbf z|\leq 1,\;z_i>0,\; z_i \in \mathbb R\},
\end{equation} 
so that, on one hand, the sum form \eqref{eqsi:temp} can be approximated with the integral 
$$\lim\limits_{n\to \infty}\left|\sum_{\mathbf z\in \Omega_n} F_n( \mathbf z) - n^{N-1} \int\limits_{\Omega_\infty} F_n(\mathbf z ) \text{d} \mathbf z\right|=0.$$
On another hand, the first fraction in Eq.~\eqref{eqA:Fn} is a properly normalised multivariate Gaussian function with mean $\mathbf z^*$  and shrinking variance $\frac{1}{n}\mathbf H^{-1}_{\mathbf z}$. This Gaussian function approaches the Dirac's delta $\delta(\mathbf z-\mathbf z^*)$ in the large $n$ limit: its variance vanishes as $O(\frac{1}{n})$ while the mean remains constant.
By combining these two observations together we obtain:
$$
\lim\limits_{n\to\infty}  \frac{w'(n)}{w'_\infty(n)}=1,
$$
where
\begin{equation}
\begin{aligned}
w'_\infty(n)  =& n^{N-1} \int\limits_{\Omega_\infty} \delta(\mathbf z-\mathbf z^*) 
\frac{\det[2 \pi  \frac{1}{n}\mathbf H^{-1}_{\mathbf z^*} ]^{1/2}}{\det[2 \pi  \boldsymbol\sigma_{\mathbf z}]^{1/2}}
e^{-g(\mathbf z^*)}
e^{-n f(\mathbf z^*)  -n \mathbf R(\mathbf z, \mathbf z^*)}=\\
&n^{N-1}  \frac{\det[2 \pi  \frac{1}{n}\mathbf H^{-1}(\mathbf z^*) ]^{1/2}}{\det[2 \pi  \boldsymbol\sigma_{\mathbf z^*}]^{1/2}}
e^{-g(\mathbf z^*)}e^{-n f(\mathbf z^*)  -n \mathbf R(\mathbf z^*, \mathbf z^*)}.
\end{aligned}
\end{equation}
Note that by the definition, the residue vanishes at the expansion point: $\mathbf R(\mathbf z^*, \mathbf z^*)=0$. Finally, by moving $2\pi$ and $\frac{1}{n}$ outside the determinants, one obtains 
\begin{equation}\label{eqsi:w_inf'}
w'_\infty(n) =C_0  n^{-1/2}e^{-nC_2 },
\end{equation}
where
$C_0=\left( \frac{ \det[  \mathbf H^{-1}_{\mathbf z^*} ]}{ 2 \pi  \det[ \boldsymbol\sigma_{\mathbf z^*}]} \right)^{1/2}\!\!e^{-g(\mathbf z^*)}$
and $C_2= f(\mathbf z^*).$

\subsection{Complete size distribution asymptote}
The characteristic function for \eqref{eqsi:D*} is given by:
$$
D_{i,j}( \boldsymbol \omega )  =  \delta_{i,j}
+\iu \frac{\partial}{\partial \omega_j} \phi_i( \boldsymbol \omega ) \phi_i^{-1}( \boldsymbol \omega ),
$$
so that the characteristic function of the full expression appearing under the sum in \eqref{eqA1:multiplex} is given by
$$
\det\left[ \boldsymbol D( \boldsymbol \omega ) \right]\phi( \boldsymbol \omega )\prod\limits_{l=1}^N \phi_l^{k_l}( \boldsymbol \omega )=\det\left[ \boldsymbol D^{'}( \boldsymbol \omega ) \right]
$$
where
$$
D^{'}_{i,j}:= \delta_{i,j}\phi^{\frac{1}{N}}(\boldsymbol \omega)
\psi^{\frac{1}{N}}(\boldsymbol \omega,\mathbf k )
+\iu \frac{\partial}{\partial \omega_j}\phi_i( \boldsymbol \omega ) \phi_i^{\frac{k_i}{N}-1 }( \boldsymbol \omega ) \phi^{\frac{1}{N}}(\boldsymbol \omega)\prod\limits_{l=\{1,...,N\}\setminus {i}} \phi_l^{\frac{k_l}{N}}( \boldsymbol \omega )
 =
$$
$$
 \delta_{i,j}
\phi^{\frac{1}{N}}(\boldsymbol \omega)\psi^{\frac{1}{N}}(\boldsymbol \omega,\mathbf k )
+\iu \frac{N}{k_i}\frac{\partial}{\partial  \omega_j} \phi_i^{\frac{k_i}{N}}( \boldsymbol \omega )\phi^{\frac{1}{N}}(\boldsymbol \omega) \prod\limits_{l=\{1,...,N\}\setminus {i}} \phi_l^{\frac{k_l}{N}}( \boldsymbol \omega ).
$$
The limiting functions for $\phi_{i}(\boldsymbol \omega)$ are given in  \eqref{eqsi:phi_inf}, so that one can write 
\begin{equation}\label{eqsi:phi_i_inf}
\phi_{i,\infty}^{\frac{k_i}{N}}( \boldsymbol \omega)=e^{\iu\, \frac{k_i}{N} \boldsymbol \mu_i^\top\boldsymbol\omega-\frac{1}{2} \boldsymbol\omega^\top \frac{k_i}{N}\boldsymbol  T_i\,\boldsymbol\omega}, 
\end{equation}
which feature the following partial derivatives 
\begin{equation}
\label{eqsi:dphi_i_inf}
\iu \frac{N}{k_i}\frac{\partial}{\partial  \omega_j} \phi_{i,\infty}^{\frac{k_i}{N}}( \boldsymbol \omega ) =-(\boldsymbol \mu_i^\top\boldsymbol -\iu \boldsymbol\omega^\top \boldsymbol  T_i\,\boldsymbol )\boldsymbol e_j \phi_{i,\infty}^{\frac{k_i}{N}}( \boldsymbol \omega),
\end{equation}
By replacing  $\phi_{i}(\boldsymbol \omega)$ and  $\iu \frac{N}{k_i}\frac{\partial}{\partial  \omega_j}\phi_{i}(\boldsymbol \omega)$ with their limiting functions \eqref{eqsi:phi_i_inf},\eqref{eqsi:dphi_i_inf} we obtain a chain of transformations:
\begin{equation*}
\begin{aligned}
D^{'}_{i,j}=& \delta_{i,j}
\phi( \boldsymbol \omega )\psi_{\infty}^{\frac{1}{N}}(\boldsymbol \omega,\mathbf k )+\iu \frac{N}{k_i}\frac{\partial}{\partial \omega_j} \phi_{i,\infty}^{\frac{k_i}{N}} ( \boldsymbol \omega)   \phi( \boldsymbol \omega ) \prod\limits_{l=\{1,...,N\}\setminus {i}} \phi_{l,\infty}^{\frac{k_l}{N}}( \boldsymbol \omega )= \\
&\delta_{i,j}\phi( \boldsymbol \omega )\psi_{\infty}^{\frac{1}{N}}(\boldsymbol \omega,\mathbf k )-(\boldsymbol \mu_i^\top\boldsymbol -\iu \boldsymbol\omega^\top \boldsymbol  T_i\,\boldsymbol )\boldsymbol e_j \phi_{i,\infty}^{\frac{k_i}{N}}( \boldsymbol \omega)\phi( \boldsymbol \omega ) \prod\limits_{l=\{1,...,N\}\setminus {i}} \phi_{l,\infty}^{\frac{k_l}{N}}( \boldsymbol \omega )=\\&
\delta_{i,j}\phi( \boldsymbol \omega )\psi_{\infty}^{\frac{1}{N}}(\boldsymbol \omega,\mathbf k )-(\boldsymbol \mu_i^\top\boldsymbol -\iu \boldsymbol\omega^\top \boldsymbol  T_i\,\boldsymbol )\boldsymbol e_j \psi_\infty( \boldsymbol \omega, \mathbf k)=\\  
&\phi^{\frac{1}{N}}( \boldsymbol \omega )\psi_{\infty}^{\frac{1}{N}}(\boldsymbol \omega,\mathbf k )[\delta_{i,j}\-(\boldsymbol \mu_i^\top\boldsymbol -\iu \boldsymbol\omega^\top \boldsymbol  T_i\,\boldsymbol )\boldsymbol e_j  ].
\end{aligned}
\end{equation*}
Determinant $\det[D']$ can be now rewritten in the matrix form:
\begin{equation}\label{eqsi:det''}
\det[D'] =\psi_\infty(\boldsymbol \omega, \mathbf k ) \det(\boldsymbol I - \boldsymbol M)  \det[D{''}],\;D_{i,j}^{''} = \delta_{i,j} +\iu \boldsymbol \omega^\top \boldsymbol t_{i,j}, 
\end{equation}
where
$$
\boldsymbol t_{i,j} =(\boldsymbol I - \boldsymbol M)^{-1} \boldsymbol  T_i\,  \boldsymbol e_j.$$
Let us expand determinant $\det[D^{'}]$ into the sum over  the set $S_N$ of all permutations of $\{1,2,\dots,N\}$:
\begin{multline}\label{eqsi:series1}
\det[D^{'}] =\psi_\infty(\boldsymbol \omega, \mathbf k ) \det(\boldsymbol I - \boldsymbol M)\sum\limits_{\sigma \in S_N}\text{sgn}(\sigma) \prod\limits_{i=1}^N (\delta_{i,\sigma_i} +  \iu \boldsymbol \omega^\top \boldsymbol t_{i,\sigma_i}) =\\ 
\psi_\infty(\boldsymbol \omega, \mathbf k )\det(\boldsymbol I - \boldsymbol M)\left(\boldsymbol c_0 + \iu \boldsymbol \omega^\top \boldsymbol c_{1,1}+ (\iu \boldsymbol \omega^\top \boldsymbol c_{2,1})(\iu  \boldsymbol \omega^\top \boldsymbol a_{2,2}) +\dots+\prod\limits_{i=1}^N \iu \boldsymbol \omega^\top \boldsymbol c_{N,i}\right), \; \boldsymbol c_{i,j}  \in \mathbb R^{N}
\end{multline}
Since the gradient of $\psi_\infty(\boldsymbol \omega, \mathbf k )$ is given by
$$
\nabla \psi_\infty(\boldsymbol \omega, \mathbf k )  =
[ \iu  (\boldsymbol M \mathbf k + \boldsymbol \mu_0 ) - \boldsymbol\sigma( \mathbf k ) \boldsymbol \omega ]  \psi_\infty(\boldsymbol \omega, \mathbf k ),
$$
one can express the $\iu \boldsymbol\omega \psi_\infty(\boldsymbol \omega, \mathbf k )$ from the latter equation as:
$$
\iu \boldsymbol\omega \psi_\infty(\boldsymbol \omega, \mathbf k )
  =
- \boldsymbol\sigma^{-1}(\mathbf k) \boldsymbol M \mathbf k \psi_\infty(\boldsymbol \omega, \mathbf k )
 - \boldsymbol\sigma^{-1}(\mathbf k)  \boldsymbol \mu_0 \psi_\infty(\boldsymbol \omega, \mathbf k )
-\iu\boldsymbol\sigma^{-1}(\mathbf k) \nabla \psi_\infty(\boldsymbol \omega, \mathbf k ),  
$$
which is the characteristic function for
\begin{equation}\label{eqsi:x}
 \boldsymbol x_n(\mathbf z) = 
 \frac{1}{n}\sigma^{-1}_{\mathbf z} ( n(\boldsymbol M- \boldsymbol I)\boldsymbol A^{-1}\mathbf z  + \boldsymbol \mu_0)  =
   -\sigma^{-1}_{\mathbf z}(\boldsymbol M- \boldsymbol I)\boldsymbol A^{-1}\mathbf z  +  \frac{1}{n} \sigma^{-1}_{\mathbf z}\boldsymbol \mu_0.
\end{equation}
Since  series \eqref{eqsi:series1} is the sum of powers of  $\iu \boldsymbol\omega \psi_\infty(\boldsymbol \omega, \mathbf k )$, this series is the characteristic function for an identical expression in which $\iu \boldsymbol\omega \psi_\infty(\boldsymbol \omega, \mathbf k )$ is substituted by \eqref{eqsi:x}:
\begin{equation}\label{eqsi:series2}
d(n,\mathbf z)=\det(\boldsymbol I - \boldsymbol M)\left(\boldsymbol c_0 + \boldsymbol x_n^\top(\mathbf z)  \boldsymbol c_{1,1}+ (\boldsymbol x_n^\top(\mathbf z)  \boldsymbol c_{2,1})(\boldsymbol x_n^\top(\mathbf z)  \boldsymbol c_{2,2}) +\dots+\prod\limits_{i=1}^N \boldsymbol x_n^\top(\mathbf z)  \boldsymbol c_{N,i}\right) F_n(\mathbf z).
\end{equation}
The latter expression is the product of $ F_n(\mathbf z)$ and a polynomial in an indeterminate $y=\frac{1}{n}$: 
\begin{equation}\label{eqsi:y}
d(n,\mathbf z) = (a_0(\mathbf z) +a_1(\mathbf z) y+a_2(\mathbf z)y^2+\dots+a_N(\mathbf z) y^{N}) F_n(\mathbf z).
\end{equation}
Instead of this polynomial, it is convenient to consider its collapsed form that we obtain by observing that the series from Eqs. \eqref{eqsi:series1} and \eqref{eqsi:series2} coincide under the substitution $\iu\boldsymbol \omega \to\boldsymbol x_n^\top(\mathbf z) ,$ therefore performing the same substitution in Eq. \eqref{eqsi:det''} leads to:
$$
d(n,\mathbf z^*) = p(n)  F_n(\mathbf z^*),
$$
where $p(n)$ is independent of $\mathbf z^*$:
\begin{equation}\label{eqsi:pn0} 
p(n) = \det( \boldsymbol M - \boldsymbol I)\det\left(\boldsymbol I+\boldsymbol Q + \frac{1}{n} \boldsymbol R \right),
\end{equation}
and
$$Q_{i,j} =\delta_{i,j}-[1,\mathbf z^*]^\top \boldsymbol K^\top \sigma^{-1}_{\mathbf z^*}\text{adj}(\boldsymbol M - \boldsymbol I)\boldsymbol  T_i\,  \boldsymbol e_j,\; 
R_{i,j} =  \boldsymbol \mu_0^\top \, \sigma^{-1}_{\mathbf z^*}\text{adj}(\boldsymbol M - \boldsymbol I)\boldsymbol  T_i\,  \boldsymbol e_j,$$
 are square matrices of size $N\times N$, and  $(\boldsymbol e_i)_j = \delta_{i,j}$ are the standard basis vectors. 
One can derive the coefficients of the expansion \eqref{eqsi:y} by applying differential operator  
$\frac{1}{k!}\frac{\partial^k }{\partial y^k}|_{y=0}$ to  \eqref{eqsi:pn0}.
For $k=0,1$ this procedure yields,
$$
a_0 = \det(\boldsymbol M - \boldsymbol I)  \det[\boldsymbol Q]=0,
$$
and
$$
a_1 = \det(\boldsymbol M - \boldsymbol I)  \text{tr}(\text{adj}(\boldsymbol Q) \boldsymbol R )>0.
$$
By taking into the account the asymptotical behaviour of $F_n(\mathbf z)$ as given by \eqref{eqsi:w_inf'}, the complete expression for the asymptote of \eqref{eqA1:multiplex} reads:
\begin{equation}
\label{eqsi:simple_asm_M1}
w_{\infty}(n)=p(n) C_0 n^{-\sfrac{1}{2} } e^{- C_2 n} =[a_1 n^{-1}+O(n^{-2})] C_0 n^{-\sfrac{1}{2} } e^{- C_2 n}\approx C_1 n^{-\sfrac{3}{2} } e^{- C_2 n},
\end{equation}
where
 $$C_1 = C_0 a_1=
\det(\boldsymbol M - \boldsymbol I)  \text{tr}(\text{adj}(\boldsymbol Q) \boldsymbol R )\left( \frac{ \det[  \mathbf H^{-1}_{\mathbf z^*} ]}{ 2 \pi  \det[ \boldsymbol\sigma_{\mathbf z^*}]} \right)^{1/2}\!\!
e^{-[1, \mathbf z^{*\top}] \boldsymbol K^\top   \boldsymbol \sigma_{\mathbf z^*} ^{-1}  \boldsymbol \mu_0}.$$
and
$$C_2 = \frac{1}{2} [1, \mathbf z^{*\top}] \boldsymbol K^\top   \boldsymbol \sigma_{\mathbf z^*} ^{-1} \boldsymbol K 
\begin{bmatrix}
1\\
\mathbf z^*
\end{bmatrix}.$$

\section{Necessary and sufficient conditions for the scale-free behaviour of $w(n)$}
In this section we show that $C_2 =0$ if and only if 
\begin{equation}\label{eqsi:criterion}
\mathbf v \in \ker( \boldsymbol M - \boldsymbol I),\; \frac{\mathbf v}{|\mathbf v|}\geq 0.
\end{equation}
According to the definition \eqref{eqsi:sigma}, $\sigma_{\mathbf z}$ is a linear combination of covariance matrices, and therefore $\sigma_{\mathbf z} ^{-1}$ is a positive definite matrix that features the Cholesky factorisation:
\begin{equation}
\label{eqsi:chol}
\boldsymbol \sigma_{\mathbf z} ^{-1} = \boldsymbol L^\top \boldsymbol L ,\; \det \boldsymbol L \neq0.
\end{equation}

Suppose $C_2 =0$. Then, according to the definition of $C_2$, $f(\mathbf z^*)=0$ and
\begin{equation}\label{eqsi:chain1}
 [1, \mathbf z^{*\top}] \boldsymbol K^\top   \boldsymbol \sigma_{\mathbf z^*} ^{-1} \boldsymbol K
\begin{bmatrix}
1\\
\mathbf z^*
\end{bmatrix}=
 [1, \mathbf z^{*\top}] \boldsymbol K^\top  \boldsymbol L^\top \boldsymbol L \boldsymbol K
\begin{bmatrix}
1\\
\mathbf z^*
\end{bmatrix}=\left(
 \boldsymbol L \boldsymbol K
\begin{bmatrix}
1\\
\mathbf z^*
\end{bmatrix} 
\right)^\top \boldsymbol L \boldsymbol K
\begin{bmatrix}
1\\
\mathbf z^*
\end{bmatrix}=0.
\end{equation}
Recall that $\boldsymbol K=(\boldsymbol M-\boldsymbol I)\boldsymbol A^{-1}$ and since $\det \boldsymbol L\neq0$ we have
$$
 \boldsymbol K \begin{bmatrix}
1\\
\mathbf z^*
\end{bmatrix} =  (\boldsymbol M - \boldsymbol I) \boldsymbol A^{-1} \begin{bmatrix}
1\\
\mathbf z^*
\end{bmatrix}=(\boldsymbol M - \boldsymbol I) \mathbf v = 0,\; \mathbf v= \mathbf A^{-1}\begin{bmatrix}
1\\
\mathbf z^*
\end{bmatrix}.$$
So that $\mathbf v \in \ker( \boldsymbol M - \boldsymbol I)$.

According to the definition of $\boldsymbol A$, $|\mathbf v| = 1-\sum\limits_{i=2}^N z^*_i +\sum\limits_{i=2}^N z^*_i=1.$
We will now demonstrate that $\mathbf v>0$.
Assume that $\mathbf z^*\notin\Omega_\infty$, then for any $ \mathbf z\in\Omega_\infty$, $f( \mathbf z )>0$, and therefore 
for arbitrary $\alpha>0$ and non-singular function $\gamma(\mathbf z)$ the following product vanish in the limit of large n, $$\lim\limits_{n \to \infty} n^\alpha \int\limits_{\Omega_n}\gamma(\mathbf z)  e^{ -\frac{1}{2}n \mathbf f( \mathbf z ) }\text{d}\mathbf z\to0,$$ 
which  contradicts our assumption that the asymptote is scale-free. 
 One therefore concludes that $\mathbf z^*\in\Omega_\infty,$ so that the set of inequalities \eqref{eqsi:Omegainfty} holds for $\mathbf z^*$. This inequalities imply:  $v_1 = 1- | \mathbf z^*|>0$ and $v_i =z_{i-1} >0, 
 $ for $i=2,3,\dots,N$,  and therefore $\frac{\mathbf v}{|\mathbf v|}=\mathbf v>0.$ This completes the proof of the forward implication.

Consider the reverse implication: suppose  $\mathbf v \in \ker (\boldsymbol M-\boldsymbol I),$ and $ \frac{\mathbf v}{|\mathbf v|}>0$.
 Without loss of generality, assume $|\mathbf v|=1$.   We will demonstrate that  vector $\mathbf z^* =(v_{2},v_{3},\dots,v_{N})^\top$
 gives minimum to $f(\mathbf z)$, and $f(\mathbf z^*)=0$.  On the one hand, we have a chain of transformations that is reverse to \eqref{eqsi:chain1}:
 $$ 0 = (\boldsymbol M-\boldsymbol I)  \mathbf v =  \boldsymbol K \begin{bmatrix}
1\\
\mathbf z^*
\end{bmatrix}
=\boldsymbol L \boldsymbol K \begin{bmatrix}
1\\
\mathbf z^*
\end{bmatrix} = 
\left(
 \boldsymbol L \boldsymbol K
\begin{bmatrix}
1\\
\mathbf z^*
\end{bmatrix} 
\right)^\top \boldsymbol L \boldsymbol K
\begin{bmatrix}
1\\
\mathbf z^*
\end{bmatrix}
=$$
$$
 [1, \mathbf z^{*\top}] \boldsymbol K^\top  \boldsymbol L^\top \boldsymbol L \boldsymbol K
\begin{bmatrix}
1\\
\mathbf z^*
\end{bmatrix}=
[1, \mathbf z^{*\top}] \boldsymbol K^\top   \boldsymbol \sigma_{\mathbf z^*} ^{-1} \boldsymbol K
\begin{bmatrix}
1\\
\mathbf z^*
\end{bmatrix}
 = f(\mathbf z^*).$$
On the other hand,  since $\mathbf f( \mathbf z) \geq 0$ for  $\mathbf z \geq 0$, $\mathbf z^*$ must also be a minimum of $\mathbf f( \mathbf z).$
 Furthermore, this minimum belongs to $\Omega_\infty.$ Indeed,  $\mathbf z^*_i =  \mathbf v_{i+1}>0$ for $i=1,\dots,N-1$, and
  $|\mathbf z^*| = \sum\limits_{i=2}^N  \mathbf v_i =|\mathbf v|-v_1<1$. The latter inequalities imply that $\mathbf z^*\in\Omega_\infty$, which 
 guarantees validity of asymptote \eqref{eqsi:simple_asm_M1}, and since $C_2=f(\mathbf z^*)=0$, this asymptote is a scale-free one.

\section{Derivations of the critical percolation probability}
Consider a percolation process on the edge-coloured network that thinners the network by randomly removing edges with probability $1-p$ (or equivalently, by keeping edges with probability $p$). Such removal of edges alters the original degree distribution. Namely, from the perspective of a randomly chosen node, each adjacent edge has equal and independent chances to be removed, so that the actual  degree distribution after percolation can be expressed by multiplying $u(\mathbf k)$ with  the binomial distribution:
\begin{equation}\label{eqsi:u_percolated}
u'(\mathbf k') = \sum\limits_{\mathbf k \geq0}\prod\limits_{i=1}^N \binom{k_i}{k_i'}p^{k_i'} (1-p)^{k_i-k_i'} u(\mathbf k).
\end{equation}
The expectations of $u'(\mathbf k')$ and those of $u(\mathbf k)$ are related:
\begin{equation}
\begin{aligned}
&\mathbb{E}[ k_i' ]        = p \mathbb{E}[ k_i ],\\
&\mathbb{E}[ k_i',k_j' ]  =p^2 \mathbb{E}[ k_i,k_j ]   + (p - p^2 ) \mathbb{E}[ k_i]  \delta_{i, j},\\
&\begin{aligned}
\mathbb{E} [ k_i', k_j', k_l' ]  =& p_ip_jp_l \mathbb{E}[ k_i, k_k, k_l]+ p_lp_i(1-p_i) \mathbb{E}[ k_i,k_l] \delta_{i, j}+\\
&p_ip_j(1-p_i) \mathbb{E}[ k_i,k_j] \delta_{i, l}+p_jp_i(1-p_j) \mathbb{E}[ k_j,k_i] \delta_{j, l}+\\
&p_i( 1 - 3 p_i + 2 p_i^2 )  \mathbb{E}\delta_{i, j}\delta_{i, l},
\end{aligned}
\end{aligned}
\end{equation}
and by plugging these substitutions into \eqref{eqsi:mu0},\eqref{eqsi:mui}, and \eqref{eqsi:T} one obtains:
 \begin{equation}
 \begin{aligned}
 &\boldsymbol \mu' = p \boldsymbol \mu,\\
& \boldsymbol M' = p \boldsymbol M,\\
  & (\boldsymbol T_i')_{j,l} = p^2(\boldsymbol T_i)_{j,l} + p (1 - p) M_{j,i} \delta_{j,l}.
 \end{aligned}
  \end{equation}
  These can be now used to compute the asymptotic properties of the percolated network.
For instance, by plugging $\boldsymbol M'$ into the criticality criterion \eqref{eqsi:criterion}, one obtains a $p$-dependant  criterion that reads: the edge-coloured network features critical percolation at $p=p_c\in(0,1]$ if there is vector $\mathbf v$, for which
\begin{equation}\label{eqsi:criterionPP}
\mathbf v \in \ker [  p_c\boldsymbol  M -\boldsymbol I],\; \frac{\mathbf v}{| \mathbf  v|} \geq 0.
\end{equation}

\section{Derivations for colour dependant percolation}
In colour-dependent percolation, the percolation probability depends on the colour of an edge. 
In this case, we consider a vector $\mathbf p =(p_1,p_2,\dots,p_N)^\top$, where $p_i$ is the probability that an edge of colour $i$ is not removed.
Colour-dependant percolation affects the degree distribution, so that after the percolation process the degree distribution becomes: 
$$u'(\mathbf k') = \sum\limits_{\mathbf k \geq0}\prod\limits_{i=1}^N \binom{k_i}{k_i'}p_i^{k_i'} (1-p_i)^{k_i-k_i'} u(\mathbf k).$$ 
Computing the expectations for the latter distribution gives:
\begin{equation}
\begin{aligned}
 &\mathbb{E}[ k_i ']     = p_i \mathbb{E}[ k_i ],\\
&\mathbb{E}[ k_i',k_j' ]   = p_i p_j \mathbb{E}[ k_i,k_j ]   + (p_i - p_i^2 ) \mathbb{E}[ k_i]  \delta_{i, j},\\
&\begin{aligned}
\mathbb{E} [ k_i, k_j, k_l ]   =& p_ip_jp_l \mathbb{E}[ k_i, k_k, k_l] +p_lp_i(1-p_i) \mathbb{E}[ k_i,k_l] \delta_{i, j}+\\
&p_ip_j(1-p_i) \mathbb{E}[ k_i,k_j] \delta_{i, l}+p_jp_i(1-p_j) \mathbb{E}[ k_j,k_i] \delta_{i, l}+\\
&p_i( 1 - 3 p_i + 2 p_i^2 )  \mathbb{E}[ k_i]\delta_{i, j}\delta_{i, l},
\end{aligned}
\end{aligned}
\end{equation}
 and plugging these into Eqs. \eqref{eqsi:mui}-\eqref{eqsi:T} allows us to express $\boldsymbol \mu, \boldsymbol M,\boldsymbol T_i $ as functions of $\mathbf p$:
\begin{equation}
\begin{aligned}
&\boldsymbol \mu' = \text{diag}\{ \mathbf p \} \boldsymbol \mu,\\
&\boldsymbol M' = \text{diag}\{ \mathbf p \}  \boldsymbol M,\\
&\boldsymbol T_i'= \text{diag}\{\mathbf p\}\boldsymbol T_i\text{diag}\{\mathbf p\}+\text{diag}\{\mathbf p\} \text{diag}\{1-\mathbf p\}\text{diag}\{ M_{1,j},\dots,M_{N,j}\}.
\end{aligned}
\end{equation}
 By plugging $\boldsymbol M'$ into Eq.~\eqref{eqsi:criterion} one obtains the criterion for criticality: edge-coloured network features a critical behaviour at percolation probability vector $0<\boldsymbol p<1$ if and only if:
 \begin{equation}\label{eqsi:criterionPP2}
\mathbf z \in \ker [\text{diag}\{ \boldsymbol p\}  \boldsymbol M-\boldsymbol P],\; \frac{\mathbf z}{| \mathbf  z|} \geq 0.
\end{equation}
 In order to recover the manifold containing all critical points $\mathbf p$ numerically, one may follow this practical procedure: \\1. for all the points from a discretised unit hypercube test if $\det(\text{diag}\{ \boldsymbol p\}  \boldsymbol M-\boldsymbol P)=0;$\\
 2. for those points that pass the first test, find  the eigenpair $(\mathbf v,0)$ of $\text{diag}\{ \boldsymbol p\}  \boldsymbol M-\boldsymbol P$ and check if $\frac{ \mathbf v}{| \mathbf v|}>0.$

\section{Derivation of colour fractions in connected components of size $n$}
For the same reason as in the conventional configuration model, every connected component of finite size $n$ has a tree-like structure, and therefore this component contains $n-1$ edges. The generalised model operates with $N$ types of edges and it is interesting to see how  these $n-1$ edges are partitioned among $N$ edge types.
 It turns out that $\mathbf z^*$, as defined in \eqref{eqsi:z*}, plays an essential role in defining this partition. Let $v_i, i=1,\dots,N$ denotes the number of edges of $i^\text{th}$ type in a component of size $n$. Since the total number of edges is $|\mathbf v|=n-1,$ one writes $v_1 = 1-\sum\limits_{i=2}^N v_i,$ and the probabilities of configurations for the rest of edge types $v_2,v_3,\dots v_N$ follow form $F_n(\mathbf z^*)$
as defined in Eq. \eqref{eqA:Fn}. In fact, since we are interested in the conditional probability given component size is $n$, it is enough to consider the first fraction appearing in \eqref{eqA:Fn}. So that the following law of large numbers holds: 
 \begin{multline}\label{eqsi:conf_dist}
\mathbb P[ \lfloor n z_1\rfloor \leq v_2 \leq\lceil n z_1\rceil  \wedge ,\dots, \wedge 
\lfloor n z_{N-1}\rfloor  \leq v_N \leq\lceil n z_{N-1}\rceil | \text{component size} = n
 ]  =\\
 \int\limits_{\lfloor n \mathbf z\rfloor}^{\lceil n \mathbf z\rceil}
 \frac{
e^{-\frac{1}{2} (\mathbf z - \mathbf z^*)^\top [\frac{1}{n} \mathbf H^{-1}_{\mathbf z^*} ]^{-1} (\mathbf z - \mathbf z^*)}
}{\det[2 \pi  \frac{1}{n}\mathbf H^{-1}_{\mathbf z^*} ]^{1/2}}
\approx
\frac{
e^{-\frac{1}{2} (\mathbf z - \mathbf z^*)^\top [\frac{1}{n} \mathbf H^{-1}_{\mathbf z^*} ]^{-1} (\mathbf z - \mathbf z^*)}
}{\det[2 \pi  \frac{1}{n}\mathbf H^{-1}_{\mathbf z^*} ]^{1/2}}.
\end{multline}
Evidently, the multivariate stochastic variable $(\frac{v_2}{n-1},\frac{v_3}{n-1},\dots, \frac{v_N}{n-1})$ is normally distributed with mean 
$\mathbf z^*$ and variance $\frac{1}{n}\mathbf H_{\mathbf z^*}, $   and  the whole vector $\frac{\mathbf v}{n-1}$ is normally distributed with  mean $$(1-|\mathbf z^*|,z^*_1,z^*_2,\dots,z^*_{N-1})$$ and covariance matrix
\begin{equation}\label{eqsi:Sigma}
\boldsymbol \Sigma = \frac{1}{n} 
\begin{bmatrix}
a& \boldsymbol b^\top\\
\boldsymbol b& \mathbf H_{\mathbf z^*} 
\end{bmatrix},
\end{equation}
where  $a=\sum\limits_{i,j=1}^{N-1} (\mathbf H_{\mathbf z^*})_{i,j}, $ and $\boldsymbol b$ is a column vector of length $N-1$,
$b_i = -\sum\limits_{j=1}^{N-1} (\mathbf H_{\mathbf z^*})_{i,j}.$

\section{Derivations for the giant component and weight-average component sizes}
Let random variable $n$ is the size of the component containing a randomly selected node. The generating function for $n$, $W(x)=\mathbb E[ x^n ],$ satisfies the system of N equations:
\begin{equation}\label{eqsi:W_and_Wi}
\begin{aligned}
W(x)&=x U[ W_1(x),\dots,W_N(x)],\\
W_1(x)&=x U_1[ W_1(x),\dots,W_N(x)],\\
\dots\\
W_N(x)&=x U_N[ W_1(x),\dots,W_N(x)].
\end{aligned}
\end{equation}
These equations constitute a generalisation of the corresponding system introduced for uncoloured networks by Newman et al.\cite{newman2001}.
In Eq. \eqref{eqsi:W_and_Wi}, $W_i(x)$ generate probabilities that a randomly selected node is connected to a component of size $m$ on either of its sides.
Since the network may contain infinitely large components, $n$ and $m$ are improper random variables. We formalise this facts by writing:
 $$
 \begin{aligned}
& g_\text{node}:=1-W(1)=1-\mathbb P[n=\infty] \leq 1,\\
& s_i:=W_i(1)=\mathbb P[n=\infty] \geq 0.
 \end{aligned}
 $$
Plugging these definitions into \eqref{eqsi:W_and_Wi} yields the following equations:
\begin{equation}\label{eqsi:g_node}
g_\text{node} = 1-\mathbb E[ \mathbf s^\mathbf k ],\; \mathbf s=(s_1,\dots,s_N)^\top,
\end{equation}
and
\begin{equation}
\label{eqsi:si}
 s_i = \frac{\mathbb E[ k_i \mathbf s^{\mathbf k-\mathbf e_i} ]}{\mathbb E[ k_i ]}, i=1,\dots,N.
\end{equation}
Here, $\mathbf e_i$ are the standard basis vectors, and vector power is evaluated element-wisely: $\mathbf s^{\mathbf k} = \prod\limits_{i=1}^N s_i^{k_i}.$
The quantities $g_\text{node}$ and $\mathbf s$ have straightforward interpretations: $g_\text{node}$, or the node size of the ginat component, is the probability that a randomly sampled node belongs to the giant component;  $ s_i $ are the probabilities that a randomly sampled $i$-coloured edge is not connected to a giant component on at least one side. So that the edge size of the giant component is written as the probability that a randomly chosen edge of colour $i$ belongs to the giant component:
$$g_i =1-s_i^2,\; i=1,\dots,N.$$
By weighting these numbers with the total fractions of coloured edges $c_i=\nicefrac{\mathbb E[k_i]}{\sum\limits_{i=1}^N \mathbb E[k_i]},$ one obtains the vector of colour fractions in the giant component, $\mathbf v^* =(v^*_1,v^*_2,\dots,v^*_N),
\; v^*_i = \frac{ g_i c_i }{\mathbf  g^\top \mathbf c}$, which is the giant-component analog of \eqref{eqsi:conf_dist}.
Since the giant component is infinite by definition, $\mathbf v^*$ is not a stochastic variable as is the case with colour fractions in  finite components \eqref{eqsi:conf_dist}. 

We will now give a quantitive estimate for the size of average finite connected component. Formally speaking, we wish to extract some properties of random variable $n'$, which is the size of randomly chosen component. 
It is important to note the subtle difference between $n$ and $n'$: they differ in the method of sampling. For the former, we first chose a node, and then take the size of the component the node belongs to, for the latter -- we chose the component itself.

Given the tools at hand, it is not easy to derive the expression for average component size $\mathbb E[n']$. 
With little efforts, however, we can  derive one for the ratio 
$$w_\text{avg}=\frac{\mathbb E[n'^2]}{\mathbb E[n']}=\frac{\mathbb E[n]}{1-g_\text{node}}=\frac{\frac{\text{d}}{\text{d}x}W(x)|_{x=1}}{1-g_\text{node}}.$$ In polymer literature, this ratio is commonly referred to as the weight-average size\cite{odian2004} and we will conveniently borrow this terminology.
The weight-average size of finite connected components $w_\text{avg}$ is found by first expressing the derivatives of $W_i(x)$ and $W(x)$ from  equation \eqref{eqsi:si}.

Let $ \mathbf y=(y_1,y_2,\dots,y_N)^\top$, $y_i=\frac{\text{d}}{\text{d}x}W_i(x)|_{x=1}$, then 
$$
\begin{aligned}
y_i =&
\left(\text{W}_1'(x) \frac{\partial }{\partial s_1}U_i[ W_1(x),\dots,W_N(x)]+\dots+\text{W}_N'(x)\frac{\partial }{\partial s_N} U_i[ W_1(x),\dots,W_N(x)]\right)\Big|_{x=1}+\\
&U_i[ W_1(1),\dots,W_N(1)],
\end{aligned}
$$
or alternatively in the matrix form: $$\mathbf y= [\boldsymbol I-\boldsymbol X(\mathbf s) ]^{-1} \mathbf s,$$
where  
 $$X_{i,j}(\mathbf s) =  \frac{\partial }{\partial s_j}U_i[ W_1(x),\dots,W_N(x)]\big|_{x=1} = \frac{ 
 \mathbb E [ ( k_i k_j - \delta_{i,j}k_i  ) \mathbf s^{ \mathbf k - \mathbf e_i - \mathbf e_j }  ] 
 }{ \mathbb E [ k_i] },\; i,j=1,\dots,N. $$
 In a similar fashion one obtains the expression for $W'(1)$:
$$
\begin{aligned}
\frac{\text{d}}{\text{d}x}W(x)|_{x=1}= &\left(\text{W}_1'(x) \frac{\partial }{\partial s_1}U[ W_1(x),\dots,W_N(x)]+\dots+\text{W}_N'(x)\frac{\partial }{\partial s_N} U[ W_1(x),\dots,W_N(x)]\right)\Big|_{x=1}+\\
&U[ W_1(1),\dots,W_N(1)]= \mathbf s^\top \boldsymbol D  \mathbf y+ 1-g_\text{node}, \;\boldsymbol D = \text{diag}\{\mathbb E[ k_i ],\dots,\mathbb E[ k_N ] \}.
\end{aligned}
$$
So that
\begin{equation}
\label{eqsi:w_avg}
w_\text{avg} =\frac{\frac{\text{d}}{\text{d}x}W(x)|_{x=1}}{1-g_\text{node}}= \frac{\mathbf s^\top \boldsymbol D[ \boldsymbol I - \boldsymbol X(\mathbf s) ]^{-1} \mathbf s }{1-g_\text{node}}+1.
\end{equation}
See also Ref \cite{kryven2018} for the derivation of the wight average size of connected components in the case of unicoloured networks.

Finally, we show how expressions \eqref{eqsi:g_node} and \eqref{eqsi:w_avg} change as a result of simple bond percolation.
In this case, the degree distribution becomes dependant on percolation probability $p$ as defined in Eq. \eqref{eqsi:u_percolated}, which induces the following transformation of the giant component size:
\begin{equation}\label{eqsi:g_node}
g_\text{node}(p) = 1-\mathbb E[ (p(\mathbf s_p-1)+1)^\mathbf k ],
\end{equation}
where
$$ (s_p)_i = \frac{\mathbb E[ k_i  (p(\mathbf s_p-1)+1)^{\mathbf k-\mathbf e_i} ]}{\mathbb E[ k_i ]}, i=1,\dots,N.$$
In a similar fashion, $w_\text{avg}$ also becomes a function of $p$:
$$
w_\text{avg}(p) = \frac{\mathbf s_p^\top \boldsymbol D[p^{-1} \boldsymbol I - \boldsymbol X(p(\mathbf s_p-1)+1) ]^{-1} \mathbf s_p }{1-g_\text{node}(p)}+1.
$$

\section{List of all phase transitions for an instance of a network with three colours  }
Consider the following degree distribution,
\begin{equation}\label{eqsi:EX2}
u(\mathbf k) =C\begin{cases}
  \text{Poiss}( k_1, 1.5 ), &  k_2,k_3 < 1;\\
  \text{Poiss}( k_2, 2.5 ), &  k_1,k_3 < 1;\\
  \text{Poiss}( k_3, 5 ),   &  k_1,k_2 < 1;\\
  \alpha,                   &  k_1,k_2,k_3 = 1.
  \end{cases}
\end{equation}
When $\alpha=0$ matix $\boldsymbol M$ has all off-diagonal elements equal to zero. Namely,
 $$\boldsymbol M = 
 \begin{bmatrix}
         2.5 &  0   &  0\\
         0   &  1.5 &  0\\
         0   &  0   &  5
\end{bmatrix}.
$$
These diagonal elements give the complete list of eigenvalues and their reciprocals $( 0.200,0.400,0.667)$ the list of all critical percolation points up to multiplicity. 
The complete list together with multiplicities is given in Table \ref{tab:pcs0}. 
\begin{table}[H]
\begin{center}
\setlength{\tabcolsep}{10pt}
\begin{tabular}{lccc}
\# & $x_S$&  $p_c$ &group \\
\hline
    1&    $(0,    0,    1)$&    0.2000 &   1\\
    2&    $(1,    0,    1)$&    0.2000 &   1\\
    3&    $(0,    1,    1)$&    0.2000 &   1\\
    4&    $(1,    1,    1)$&    0.2000 &   1\\
    5&    $(1,    0,    0)$&    0.4000 &   2\\
    6&    $(1,    1,    0)$&    0.4000 &   2\\
    7&    $(1,    0,    1)$&    0.4000 &   2\\
    8&    $(1,    1,    1)$&    0.4000 &   2\\
    9&    $(0,    1,    0)$&    0.6667 &   3\\
   10&    $(1,    1,    0)$&    0.6667 &   3\\
   11&    $(0,    1,    1)$&    0.6667 &   3\\
   12&    $(1,    1,    1)$&    0.6667 &   3
\end{tabular}
\end{center}
\caption{List of all phase transitions in the three-colour network defined by Eq.~\eqref{eqsi:EX2}, $\alpha=0$.}
\label{tab:pcs0}
\end{table}%

\bibliographystyle{apsrev4-1}
\bibliography{literature}
\end{document}